\documentclass[prx,twocolumn,english,superscriptaddress,floatfix,longbibliography]{revtex4-2}

\usepackage[utf8]{inputenc}

\usepackage{algorithm}
\usepackage{algpseudocode}
\usepackage{graphicx}
\usepackage{amsmath,amssymb,amsthm,mathrsfs,amsfonts,dsfont,mathtools}
\usepackage{physics}
\usepackage{tikz}

\usepackage[caption=false]{subfig}
\usepackage{booktabs}

\usepackage{colortbl}

\usepackage{xcolor}
\definecolor{erblue}{HTML}{0082F0}
\definecolor{erorange}{HTML}{FF8C0A}
\definecolor{ergreen}{HTML}{0FC373}
\definecolor{erpurple}{HTML}{AF78D2}
\definecolor{eryellow}{HTML}{FAD22D}
\definecolor{erred}{HTML}{FF3232}
\definecolor{grey}{HTML}{D3D3D3}

\usepackage{hyperref} 
\hypersetup{
colorlinks=true,
linkcolor=blue,
urlcolor=blue,
citecolor=erred
}

\definecolor{bondiblue}{rgb}{0.0, 0.58, 0.71}

\usepackage{soul}
\sethlcolor{yellow}

\usepackage{amsthm}
\theoremstyle{plain}
\newtheorem{definition}{Definition}

\newtheorem{problem}{Problem}
\newtheorem*{conjecture*}{Informal Conjecture}

\begin{document}
\title{Problem-informed Graphical Quantum Generative Learning}

\author{Bence Bak\'o}
\thanks{bako.bence@wigner.hun-ren.hu}
\affiliation{Eötvös Loránd University, Budapest, Hungary}
\affiliation{HUN-REN Wigner Research Centre for Physics, Budapest, Hungary}

\author{D\'aniel T. R. Nagy}
\affiliation{Eötvös Loránd University, Budapest, Hungary}
\affiliation{HUN-REN Wigner Research Centre for Physics, Budapest, Hungary}

\author{P\'eter H\'aga}
\affiliation{Ericsson Research, Budapest, Hungary}

\author{Zs\'ofia Kallus}
\affiliation{Ericsson Research, Budapest, Hungary}

\author{Zolt\'an Zimbor\'as}
\thanks{zoltan.zimboras@helsinki.fi}
\affiliation{Eötvös Loránd University, Budapest, Hungary}
\affiliation{HUN-REN Wigner Research Centre for Physics, Budapest, Hungary}
\affiliation{Algorithmiq Ltd, Kanavakatu 3C, Helsinki, 00160, Finland}
\affiliation{Department of Physics, University of Helsinki, Helsinki, Finland}

\begin{abstract}
Leveraging the intrinsic probabilistic nature of quantum systems, generative quantum machine learning (QML) offers the potential to outperform classical learning models. Current generative QML algorithms mostly rely on general-purpose models that, while being very expressive, face several training challenges. One potential way to address these setbacks is by constructing problem-informed models that are capable of more efficient training on structured problems. In particular, probabilistic graphical models provide a flexible framework for representing structure in generative learning problems and can thus be exploited to incorporate inductive bias into QML algorithms. In this work, we propose a problem-informed quantum circuit Born machine Ansatz for learning the joint probability distribution of random variables, with independence relations efficiently represented by a Markov network (MN). We further demonstrate the applicability of the MN framework in constructing generative learning benchmarks and compare our model's performance to previous designs, showing that it outperforms problem-agnostic circuits. Based on a preliminary analysis of trainability, we narrow down the class of MNs to those exhibiting favourable trainability properties. Finally, we discuss the potential of our model to offer quantum advantage in the context of generative learning, and present a practical scenario demonstrating the scalability of the approach.

\end{abstract}

\maketitle

\section{Introduction}

\begin{figure*}[t]
    \includegraphics[width=.9\textwidth]{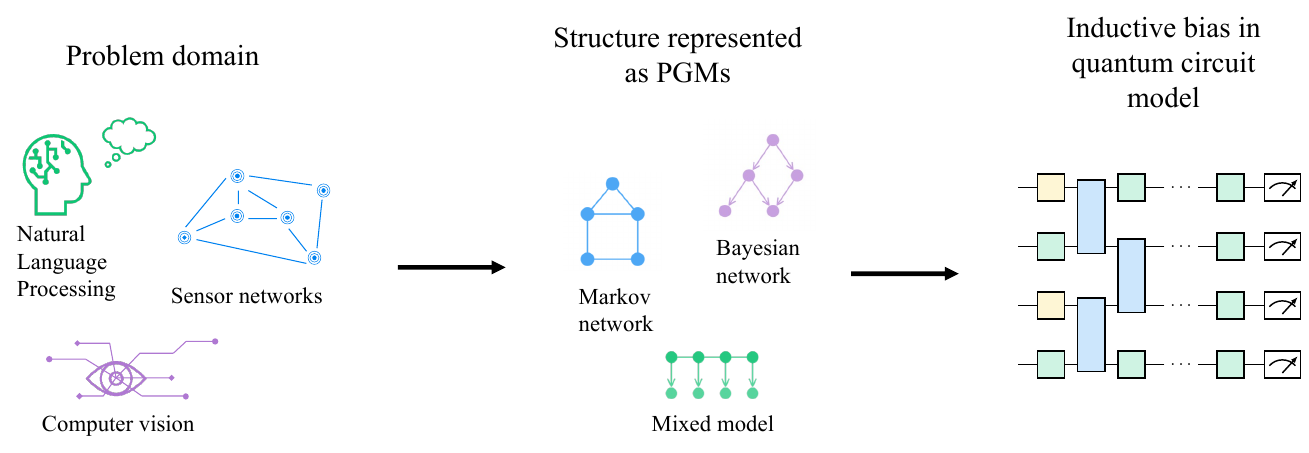}
    \caption{\textbf{Framework for using probabilistic graphical models to introduce inductive bias in quantum generative learning models.} Most generative learning problems can be (re)formulated with binary random variables. If we have enough knowledge about the problem, we can identify structure in the form of PGMs. Having a PGM, we can construct and train the corresponding quantum model that later can be used to draw samples from the learned probability distribution.}
    \label{fig:architecture}
\end{figure*}

In recent years, the field of machine learning (ML) experienced unprecedented growth, giving rise to a wide variety of models and algorithms with the potential to revolutionize several fields~\cite{sarker2021machine}. 
Generative learning is a powerful paradigm in ML that aims to capture the underlying distribution of data in order to generate realistic samples ~\cite{bond2021deep}.
Quantum ML (QML) has emerged as a promising intersection of quantum computing and ML in the pursuit of practical quantum advantage~\cite{biamonte2017quantum, schuld2014quest, liu2021rigorous, xiao2023practical, xu2025toward}.

Quantum resources, by leveraging phenomena such as quantum interference and entanglement, can be used to efficiently draw samples from probability distributions of high complexity~\cite{terhal2002adaptive, aaronson2011computational, farhi2016quantum, arute2019quantum, madsen2022quantum}. 
This makes generative QML a natural pathway towards harnessing the potential of quantum computers. 
Recent advances in this field led to the adaptation of several successful classical generative models~\cite{tian2023recent}.
The resulting architectures include quantum circuit Born machines (QCBMs)~\cite{benedetti2019generative, liu2018differentiable, coyle2020born}, quantum generative adversarial networks (QGANs)~\cite{lloyd2018quantum, dallaire2018quantum} and quantum Boltzmann machines (QBMs)~\cite{amin2018quantum, zoufal2021variational}. 

While their high expressivity makes general-purpose QML models very powerful, they also pose several challenges. 
Contrary to classical neural networks, variational quantum circuits are much more affected by trainability issues, such as barren plateaus and poor local minima~\cite{mcclean2018barren, holmes2022connecting, ragone2024lie, fontana2024characterizing, anschuetz2022quantum}. 
Furthermore, the no-free-lunch theorem~\cite{ho2002simple} also translates to QML, suggesting that these problem-agnostic models, like hardware-efficient Ans\"atze, have poor average performance~\cite{poland2020no, sharma2022reformulation}.
The reason behind these barriers can be seen as the lack of sufficient inductive bias, i.e., assumptions about the data that could be encoded into the learning framework. 
Consequently, a potential way of dealing with them is by constructing problem-informed models that can be trained more efficiently for structured problems~\cite{larocca2022group, meyer2023exploiting, zheng2023speeding, bowles2023contextuality, ding2024quantum}. Furthermore, while the mechanism of mapping classical graph structures to quantum circuits is a standard paradigm in quantum optimization~\cite{farhi2014quantum}, generative QML has largely lacked a unifying framework for this structured approach.

Probabilistic graphical models (PGMs) provide a mathematical framework for representing structure in generative learning problems defined over random variables~\cite{koller2009probabilistic} and, as such, can be exploited to construct problem-informed QML models, as depicted in Fig.~\ref{fig:architecture}. 
The two main classes of PGMs are Bayesian networks (BNs) and Markov networks (MNs). 
BNs provide a highly interpretable approach to graphical modelling by using a directed graph, leading to numerous real-world applications across many disciplines~\cite{heckerman1995real}. 
However, in certain domains, such as those involving spatial or relational data, MNs provide a more natural representation, since they do not define the orientation of the graph edges~\cite{murphy2012machine}.

While there are several excellent works concerning the quantum circuit implementation of BNs~\cite{low2014quantum, borujeni2021quantum, gao2022enhancing}, MNs are not well-studied in the context of QML. 
In this work, we investigate the applicability of the framework provided by MNs to generative QML with classical data.
We propose a problem-informed model, called Quantum Circuit Markov Random Field (QCMRF), that aims to learn the distribution over binary random variables, where the independence relations are efficiently represented by a MN.
The core of our model is a parametrized diagonal unitary operator that takes inspiration from the matrix representation of the clique structure.
The resulting procedure constructs a heuristic Ansatz that is not a strict quantum circuit implementation of the corresponding MN, but, as we demonstrate in Appendix~\ref{appendix:qcmrf-ansatz}, it is supported by its matrix representation.
This construction assumes access to a MN that provides a good approximation of the target problem, however, in several application domains, such as image processing, such approximations can be easily constructed using basic domain knowledge, e.g., grid topologies based on pixel proximity~\cite{besag1986statistical}.
While we study our model in the paradigmatic framework of QCBMs, the Ansatz design process is applicable to other well-established generative QML models, such as QGANs or quantum variational autoencoders. We numerically demonstrate this applicability as a modular building block in Appendix~\ref{app:gan}.

As opposed to previous problem-agnostic QCBMs, this Ansatz can capture higher-order correlations between the corresponding random variables and potentially reduce the number of trainable parameters, while also increasing performance. 
Throughout the paper, by higher-order correlations, we refer to correlations that cannot be described by fewer-body (e.g., pairwise) relationships.
We argue that this new model class has the potential to demonstrate quantum advantage, since it contains the class of QAOA circuits~\cite{farhi2014quantum} that were shown to produce classically hard probability distributions~\cite{farhi2016quantum, krovi2022average}.
Besides model design, we also show the potential of the PGM framework to construct benchmarks for generative QML models, where the problem complexity can be fine-tuned in multiple factors. 
We perform numerical experiments based on this benchmark proposal against both problem-agnostic and BN-based QML models. We also present a preliminary analysis of trainability and define a class of efficient graphical representations that have higher potential in this context. 

It is important to note that, while we concentrate on the significance of problem-specific model construction in the context of QML, this trend is also present in classical ML, where much larger models can be implemented~\cite{von2021informed}. This further illustrates the power of problem-informed approaches, relevant not only for near-term devices, but potentially for the large-scale fault-tolerant quantum computing as well.
To demonstrate the scalability of our Ansatz design framework, we consider a large-scale training of an instantaneous quantum polynomial-time (IQP) QCBM~\cite{recio2025train}, where the structure of the diagonal quantum circuit follows the QCMRF construction using MNs relevant in the image processing domain. Remarkably, while these MNs are not guaranteed to capture the exact correlation structure of the dataset, they can significantly reduce the required number of trainable parameters, while maintaining performance.

The structure of this paper is as follows: Sec.~\ref{sec:background} offers a review of relevant concepts regarding PGMs and generative QML. In Sec.~\ref{sec:results}, we present our generative QML model and benchmark proposal, along with numerical experiments. Finally, in Sec.~\ref{sec:conclusion}, we discuss potential future directions.

\section{Background \& Notation}
\label{sec:background}

\subsection{Probabilistic Graphical Modelling}
\label{subsec:pgms}

Explicitly encoding the probabilities of each assignment in a high-dimensional state-space is infeasible, as it scales exponentially with the number of random variables. In a space of only ten binary variables, we would need $2^{10} -1 = 1023$ numbers to represent a probability distribution. PGMs were developed to tackle this problem by using a graph representation to compactly encode a complex distribution of interacting random variables.
These graphs effectively capture the independence relations between variables and enable to split the joint probability distribution into smaller factors, each over a smaller subspace. Furthermore, this framework is also useful for inference and learning tasks. 

Here we give a brief description of the mathematical framework of the two main families of PGMs and their connections.
Since any higher-order PGM can be embedded into PGMs over binary random variables, we focus our attention on the latter. For an extensive study of PGMs, we refer the reader to Ref.~\cite{koller2009probabilistic}.

\subsubsection{Bayesian networks}
BNs use directed acyclic graphs to represent the conditional dependencies between random variables. In these models, a variable is independent of all other variables given its parents in the graph. Consequently, the factors of the joint probability distribution can be interpreted simply as conditional probabilities. 

\begin{definition}[Bayesian network factorization]
A distribution $P_{\mathcal{B}}$ over the space of $n$ random variables $\mathbf{X} = \{X_1, \dots, X_n\}$ factorizes according to a Bayesian network $\mathcal{B} = (\mathcal{G}, P)$, if $P_{\mathcal{B}}$ can be expressed as a product
$$P(X_1,\dots, X_n) = \prod\limits_{i=1}^{n} P(X_i| Pa^{\mathcal{G}}_{X_i}),$$
where $Pa^{\mathcal{G}}_{X_i}$ denotes the parents of the node associated to variable $X_i$ in graph $\mathcal{G}$.
\label{def:bn}
\end{definition}

As an example, in the BN presented in Fig.~\ref{subfig:bayes}, the set of independencies can be written as: $(B \perp C\;|\;A)$, $(D \perp A\;|\;B, C)$, $(E \perp C, D\;|\;B)$. As per Def.~\ref{def:bn}, the joint probability distribution is $P(A,B,C,D,E) = P(A)P(B|A)P(C|A)P(D|B,C)P(E|B)$.

\subsubsection{Markov networks}

Markov networks, Markov random fields or undirected graphical models are defined over general undirected graphs to represent a set of random variables having the Markov property. 
In general, the global Markov property applies, which states that any two subsets of variables are conditionally independent given a separating subset. 
As opposed to BNs, where the factors are straightforward to comprehend, here they cannot be interpreted directly. 
However, we can view a factor as describing ``compatibilities'' between different values of the variables in the corresponding subset. 
A factor here describes a general purpose function $\phi: \mathrm{Val}(\mathbf{D}) \rightarrow \mathbb{R}$, where $\mathrm{Val}(\mathbf{D})$ denotes all possible joint states of a set of random variables $\mathbf{D}$.
Each factor corresponds to a clique in the graph, however, the usual graphical representation does not make it clear, whether the joint probability distribution only factorizes according to the maximal cliques or also the subsets thereof. However, the maximal clique factorization is the most general for a given graph, therefore, we work with this assumption.

\begin{figure}[t]
\centering
\subfloat[0.23\textwidth][Bayesian\ network]{
\label{subfig:bayes}
\includegraphics[width=0.42\columnwidth]{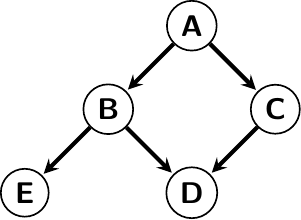}
}
\quad\quad
\subfloat[0.23\textwidth][Markov network]{
\label{subfig:mrf}
\includegraphics[width=0.3\columnwidth]{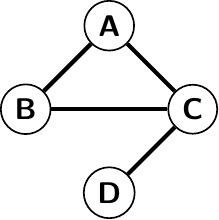}
\hspace*{4pt}
}
 \caption{\textbf{Small examples of the two main classes of probabilistic graphical models.} The nodes represent random variables, the edges represent some dependence between them. Bayesian networks have directed edges, denoted by arrows, Markov networks have undirected edges.}
\end{figure}

\begin{definition}[Markov network factorization]
    We say that a distribution $P_{\Phi}$ with $\Phi = \{\phi_1(\mathbf{D}_1), \dots, \phi_K(\mathbf{D}_K)\}$ factorizes over a Markov network $\mathcal{H}$ if each $\mathbf{D}_k$ $(k=1, \dots, K)$ is a complete subgraph of $\mathcal{H}$ and $P_{\Phi}$ is a Gibbs distribution parametrized by these factors as follows:
    \begin{equation*}
    \begin{split}
         P_{\Phi}(X_1, \dots, X_N) & = \frac{1}{Z} \tilde{P}_{\Phi}(X_1, \dots, X_N) \\
         & =  \frac{1}{Z} \phi_1(\mathbf{D}_1) \times \phi_2(\mathbf{D}_2) \times \dots \times \phi_K(\mathbf{D}_K),
    \end{split}
    \end{equation*}
    where $\times$ denotes the factor product, $\tilde{P}_{\Phi}$ is the unnormalized measure and $$Z = \sum\limits_{X_1, \dots, X_n} \tilde{P}_{\Phi}(X_1, \dots X_n)$$
    is the normalizing constant or partition function.
    \label{def:mn}
\end{definition}

For the MN in Fig.~\ref{subfig:mrf}, with independence relations $(A \perp D \; | \; C)$, $(B \perp D \; | \; C)$ and assuming maximal clique factorization, the joint probability distribution, according to Def.~\ref{def:mn}, takes the form $P(A,B,C,D) = \frac{1}{Z} \phi_1(A,B,C) \phi_2(C,D)$. Here $\phi_1(A,B,C)$ and $\phi_2(C,D)$ denote the two maximal cliques of the graph.

In this picture, the factors are encoded as complete tables, however, for positive factor values, there exists an alternative parametrization that connects MNs to energy-based models~\cite{song2021train}. In this case, the joint probability distribution can be written as
\begin{equation}
    P_{\Phi}(X_1, \dots, X_N) = \exp{\left[-\sum\limits_{i=1}^{m} \epsilon_i(\mathbf{D}_i)\right]},
\end{equation}
where $\epsilon_i(\mathbf{D}_i) = -\ln{\phi_i(\mathbf{D}_i)}$ is called the energy function. This parametrization of MNs is usually referred to as the log-linear model. We use this representation as inspiration for constructing our MN-based QML model.

\subsubsection{Connection between BNs and MNs}

Bayesian and Markov networks are incomparable in terms of independence relations they capture. 
However, we can convert one type into the other that can represent the same probability distribution by potentially introducing new edges between the nodes of the graph.

The transformation from BNs to MNs, called moralization, follows a simple rule: if two nodes are connected by a directed edge in the BN graph, or they are both parents of at least one common node, then they are connected by an undirected edge in the MN graph. Having the undirected graph, we can assign a factor to each resulting clique, to obtain a MN. 
This procedure usually leads to a PGM with more parameters that can lead to a longer training.
Directed graphs, in which parent nodes do not share common children, are called moral. Consequently, BNs that are defined over moral graphs, can be converted to MNs without introducing additional edges and parameters. 

Turning MNs into BNs is a more difficult task, both conceptually and computationally. First, the undirected graph needs to be triangulated, meaning that we introduce chords in cycles of $4$ or more and repeat this process until there are no such cycles left. Chords connect nodes that are in the cycle, but are not already connected. This transformation usually leads to the introduction of a much larger number of edges than in the previous case.  Finally, the undirected edges have to be turned into directed ones in an acyclic manner.
Since chordal graphs are also moral, PGMs that are defined over chordal graphs represent a class of graphical models that can be treated as either BNs or MNs and the conversion between the two is straightforward. 

\subsubsection{Generative learning in PGMs}
The two main learning tasks in the context of PGMs are structure learning and parameter estimation~\cite{koller2009probabilistic}. 
In this work, we restrict our attention to a version of parameter estimation in a generative learning setting. We require our model to learn how to sample the unknown probability distribution that the corresponding PGM induces. 
We give a formal description of this problem for MNs, but it can be formulated for BNs analogously to Prob.~\ref{prob:mrf-learning}. Throughout this work, we use the phrases distribution learning and generative learning interchangeably.

\begin{problem}[Distribution Learning in MNs]
    Given the graph structure and the clique factorization of a  Markov network $\mathcal{H}$ with an unknown joint probability distribution $P^*$, a dataset $\mathcal{D}$ sampled from $P^*$ and $\varepsilon, \delta \in (0,1)$, output with probability at least $1-\delta$ a representation of a distribution $P_{\mathcal{M}}$ satisfying $d(P^*, P_{\mathcal{M}})\leq \varepsilon$.
\label{prob:mrf-learning}
\end{problem}

In the above problem formulation, $d(\cdot,\cdot)$ refers to the distance between the two distributions. In this work, we focus on the total variation (TV) distance defined as
\begin{equation}
    \mathrm{TV}(P^*, P_{\mathcal{M}}) = \frac{1}{2} \sum\limits_{x\in \{0,1\}^n} |P^*(x)-P_{\mathcal{M}}(x)|.
    \label{eq:tv}
\end{equation}

In MNs, the use of a normalizing constant couples the parameters across the whole network, which prevents us from decomposing the problem and estimating local groups of parameters separately. One of the computational ramifications of this global coupling is that not even the parameter estimation with complete data can be solved in closed form (except for some special cases that can essentially be reformulated as BNs~\cite{gogate2010learning}). This makes the use of iterative methods, such as gradient descent, unavoidable. Luckily, the objective function is convex, meaning that these methods are guaranteed to converge. However, they come with the disadvantage of having to run inference in each step to calculate the gradients and inference in MNs is \#P-complete in general, which makes the distribution learning task formulated above fairly expensive or even intractable classically~\cite{roth1996hardness}.

\subsection{Generative QML Models}
\subsubsection{General framework of QCBMs}
\label{sec:qcbm}
QCBMs, introduced in Ref.~\cite{benedetti2019generative}, are paradigmatic generative QML models that naturally inherit the Born rule and thus can be used to generate tunable and discrete probability distributions that approximate a target distribution. In this section, we review the general components of variational quantum algorithms in the context of QCBMs.

One of the key building blocks is the Ansatz. The Ansatz refers to a parametrized family of quantum circuits used as a hypothesis or approximation for solving the problem in question by iteratively optimizing the parameters to match the desired target behaviour or to minimize a cost function. 
We can differentiate problem-informed and problem-agnostic Ans{\"a}tze.

\begin{figure}
    \includegraphics[width=.4\textwidth]{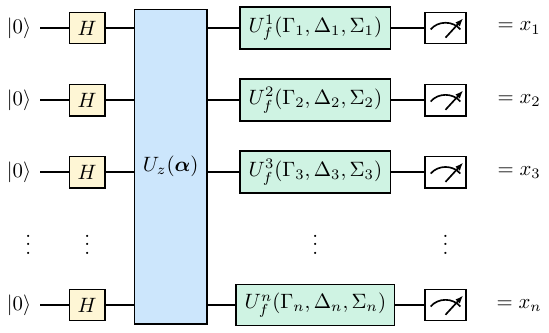}
    \caption{\label{fig:qcibm}\textbf{General structure of the quantum circuit Ising Born machine, adapted from~\cite{coyle2020born}.} The circuit starts with Hadamard gates on each qubit, then a parametrized $n$-qubit unitary is applied with tunable parameters $\boldsymbol{\alpha}$, finally, a general parametrized $U_f$ gate acts on each qubit separately. The measurement is taken in the computational basis.}
\end{figure}
The quantum circuit Ising Born machine (QCIBM), introduced by Coyle et al.~in~\cite{coyle2020born}, in its most general form, is a problem-agnostic Ansatz. The structure of the QCIBM circuit is depicted in Fig.~\ref{fig:qcibm}, where the corresponding unitaries can be written as

\begin{equation}
    U_z(\boldsymbol{\alpha}) = \prod\limits_{j}{} U_z(\alpha_j, S_j) =  \prod\limits_{j} \exp{\left(i\alpha_j \bigotimes\limits_{k \in S_j}Z_k\right)},
    \label{eq:qcibm-uz}
\end{equation}

\begin{equation}
    U_f(\mathbf{\Gamma}, \mathbf{\Delta}, \mathbf{\Sigma}) =  \exp{\left(i\sum\limits_{k=1}^{n} (\Gamma_k X_k + \Delta_k Y_k + \Sigma_k Z_k)\right)}.
    \label{eq:qcibm-uf}
\end{equation}

Here each $S_j$ indicates a subset of qubits and the $X, Y, Z$ operators are the Pauli-matrices. The $U_f$ operators can also be thought of as a ``parametrized measurement", or letting the measurements be in any local basis.
The authors restrict the Hamiltonian that generates $U_z$ to only contain one- and two-body terms, since ``only single and two-qubit gates are required for universal quantum computation" and they consider each qubit-pair (all-to-all connectivity).

The goal of generative learning is to draw samples from a model probability distribution $P_{\boldsymbol{\theta}}$ that is sufficiently close to the target distribution $P^*$, while only having access to a finite number of samples from $P^*$.
In the case of QCBMs, the model probability distribution is approximated by repeatedly running the circuit and measuring an observable each time. The final measurements are performed in the computational basis on each qubit, producing a binary string of variables $x_i \in \{0,1\}$ from the distribution
\begin{equation}
P_{\boldsymbol{\theta}}(\mathbf{x}) = |\langle \mathbf{x}| U(\boldsymbol{\theta})|0\rangle^{\otimes n}|^2,
\end{equation}
where the $\boldsymbol{\theta}$ vector contains all circuit parameters.

There are multiple ways to characterize the distance of two distributions. The TV distance, shown in \eqref{eq:tv}, is a good benchmark, but it is not feasible as a cost function, as it uses the probabilities explicitly. While in principle, one can use an approximation of the probability distributions based on a finite number of samples, the required number is exponential in general.
One of the most common metrics used as a cost function in generative modelling is the Kullback-Leibler (KL) divergence
\begin{equation}
    D_{KL} (P_{\boldsymbol{\theta}}, P^*) = \sum_x P^*(x) \log\left(\frac{P^*(x)}{P_{\boldsymbol{\theta}}(x)}\right),
    \label{eq:kldiv}
\end{equation}
where in practice $P(x)$ is replaced by an infinitesimal constant for $0$ probabilities.

Unfortunately, this function is also unsuitable for data-driven training, as it requires a large number of training samples, and it was shown not to be trainable for larger-scale QCBMs~\cite{rudolph2024trainability}. In Ref.~\cite{liu2018differentiable} a more efficient cost function was introduced for gradient-based training of QCBMs. The idea was to compare the distance of the samples drawn from the target and the model distribution in a kernel feature space. This loss function is called the squared maximum mean discrepancy (MMD):
\begin{equation}
\begin{split}
\mathcal{L}_{\text{MMD}}(P_{\boldsymbol{\theta}}, P^*) & = \left\| \sum_x P_{\boldsymbol{\theta}}(x) \phi(x) - \sum_x P^*(x) \phi(x) \right\|^2 \\
& = \underset{x \sim P_{\boldsymbol{\theta}}, y \sim P_{\boldsymbol{\theta}}}{\mathbb{E}} [K(x,y)] + \underset{x \sim P^*, y \sim P^*}{\mathbb{E}} [K(x,y)] \\
& - 2 \underset{x \sim P_{\boldsymbol{\theta}}, y \sim P^*}{\mathbb{E}} [K(x,y)],
\end{split}
\label{eq:mmd}
\end{equation}
where the $\phi$ function maps $x$ into a higher dimensional space and by definition $K(x,y) = \phi(x)^T\phi(x)$ is the kernel function, for which they used a Gaussian kernel
\begin{equation}
    K(x,y) = \exp\left( -\frac{1}{2\sigma}|x-y|^2\right).
    \label{eq:kernel}
\end{equation}

Having a family of parametrized circuits and a cost function to characterize the distance between the circuit output and the target distribution, the final building block is the optimizer that adjusts the circuit parameters given the cost function. Both gradient-based and gradient-free optimizers have been used to train QCBMs~\cite{liu2018differentiable, coyle2021quantum}. In our numerical experiments, we only consider gradient-based optimization methods, but since our approach concentrates on Ansatz-design, it is compatible with other optimizers as well.

\subsubsection{Existing quantum circuit adaptations of PGMs}

Liu and Wang~\cite{liu2018differentiable} made the first explicit connection between the paradigmatic QCBM model and PGMs. They propose a framework, in which they first construct the Chow-Liu tree of a dataset~\cite{chow1968approximating} based on the mutual information between all pairs of bits from the training samples. Having this tree graph, they propose a QCBM Ansatz, in which, the connectivity pattern of the $CNOT$ gates respects the graph structure. The Chow-Liu tree offers an effective approach for creating a second-order product approximation of a joint probability distribution. The corresponding graph represents a BN that can also be regarded as a pairwise MN, but being a second-order approximation, it fails to detect higher-order correlations. 
Another explicit connection between BNs and QCBMs was formulated in Ref.~\cite{benedetti2021variational}, where the authors proposed a framework that utilizes QCBMs for variational inference in PGMs.

Besides the variational framework of general-purpose QCBMs, there have been several attempts to exactly implement PGMs on a quantum computer. 
BNs have an equivalent formulation in the computational basis measurements of a class of quantum circuits known as Bayesian quantum circuits (BQCs)~\cite{low2014quantum}. These are defined such that the probability distribution they sample from, by measuring the given qubits in the computational basis, corresponds to the distribution defined by the corresponding BN. BQCs are implemented with uniformly controlled gates that can be decomposed into one-qubit rotations and $CNOT$ gates~\cite{borujeni2021quantum, PhysRevA.71.052330}. By definition, these circuits obey certain rules in accordance with the directed, acyclic nature of the underlying graph.
 
 In Ref.~\cite{gao2022enhancing} the authors introduced a minimal extension to BQCs and presented unconditional proof of separation in the expressive power of BNs and the corresponding basis-enhanced BQCs (BBQCs). They showed that by letting the final measurement be in any local basis, this separation appears that can be associated with quantum nonlocality and contextuality. They also pointed out that both BQCs and their basis-enhanced versions can be efficiently simulated with classical tensor network methods, when the graphs are sparse enough.
 
 The literature on the quantum circuit implementation of Markov networks is more scarce. In Ref.~\cite{piatkowski2024quantum} the authors identified a novel embedding technique of MNs into unitary operators that relies on their log-linear representation. They construct a Hamiltonian composed of Pauli-$Z$ terms and give a quantum algorithm that implements the exponential of this Hamiltonian, meaning that measuring the output qubits of the corresponding quantum circuit is equivalent to sampling the corresponding MN. 
The circuit that implements this exponentiation, uses a special point-wise polynomial approximation and real part extraction that might fail. For this reason, they have to measure ancillary qubits in order to determine whether this extraction was successful and potentially start everything over. Consequently, the success probability decreases exponentially with the number of maximal cliques. This can of course be amplified with quantum singular value transformation~\cite{gilyen2019quantum}, but that further increases the required resources. 
Since Boltzmann machines form a subclass of pairwise MNs, their quantum circuit implementations can also be regarded as an adaptation of PGMs~\cite{zoufal2021variational}. However, these models are quite restricted compared to general MNs, since they only consider pairwise correlations, usually in a bipartite manner. 

\section{Quantum Circuit Markov Random Fields}
\label{sec:results}
In this section, we present our results, starting with the definition of our QML model, proposed for generative learning in MNs. We then introduce our novel benchmark proposal and compare our model to both problem-agnostic QCIBMs and BBQCs through a series of numerical experiments. As a preliminary analysis of trainability, we investigate the scaling of the cost function variance for different types of graphs. Finally, we present our argument for a potential quantum advantage of our MN-based model class and demonstrate the scalability of the approach.

\subsection{From Graphical Representation to Variational Ansatz}
\label{sec:qcmrf}
We propose a QCBM Ansatz for distribution learning in MNs, as described in Prob.~\ref{prob:mrf-learning}.
We start by constructing a parametrized many-body Ising Hamiltonian that is inspired by the log-linear model of MNs, and consequently depends on the clique structure of the MN $\mathcal{H}$. This Hamiltonian takes the form
\begin{equation}
    H' (\boldsymbol{\beta})= \sum_{C \in \mathcal{C}_{\mathcal{H}}} \bigotimes_{v \in C} \beta_{C, v} (I+Z_v),
\end{equation}
where $Z_v$ is the Pauli-$Z$ operator acting on qubit $v$, $\mathcal{C}_{\mathcal{H}}$ refers to the set of cliques and $\boldsymbol{\beta}$ is the set of parameters. Please note that the $\boldsymbol{\beta}$ parameters are only used to enable a compact representation, but ultimately do not reflect the number of free parameters in the final model.
Usually some of the MN cliques overlap in non-zero subsets, thus there will be reoccurring terms. 
Since all terms commute, we can reparametrize the Hamiltonian such that each term only appears once with a distinct coefficient (and identities are excluded): 
$H'(\boldsymbol{\beta}) \rightarrow H(\boldsymbol{\alpha})$. Also note that in this new parametrization, each element of $\boldsymbol{\alpha}$ is independent, in the most general case, and $|\boldsymbol \alpha| \geq |\boldsymbol \beta|$.

Having formulated this parametrized Hamiltonian that encodes the structure of the problem, we consider the unitary it generates as
\begin{equation}
 U_Z(\boldsymbol{\alpha}) = e^{-i H(\boldsymbol{\alpha})},
\end{equation}
and implement a model similar to QCIBMs, defined in Sec.~\ref{sec:qcbm} and Fig.~\ref{fig:qcibm}. We call these problem-informed QCBM models QCMRFs. For a more detailed model construction process, please refer to Appendix~\ref{appendix:qcmrf-ansatz}, where we rely on the representation of probability distributions as diagonal matrices and use the Pauli expansion of the projectors. Furthermore, we also show that the output probability distribution of the QCMRF can be factorized according to the same MN that was used in its construction.

As an example, the MN shown in Fig.~\ref{subfig:mrf} defines a $3$-local Hamiltonian of the following form: $H(\boldsymbol{\alpha}) = \alpha_1 Z_AZ_BZ_C + \alpha_2 Z_AZ_B + \alpha_3 Z_BZ_C  + \alpha_4 Z_AZ_C + \alpha_5 Z_CZ_D + \alpha_6 Z_A + \alpha_7 Z_B + \alpha_8 Z_C + \alpha_9 Z_D$. 

Alternatively, one can also limit the locality of the Hamiltonian to get shallower circuits that in turn can be worse at capturing higher-order correlations. This is ultimately equivalent to considering smaller cliques instead of the maximal clique factorization. 
The circuit implementation details are discussed in Appendix~\ref{appendix:qcmrf-implem}.

\subsection{Benchmark Proposal}

Benchmarking generative QML models often relies on generic probability distributions, such as the bars and stripes dataset, or some Hamming weight specific target distribution~\cite{benedetti2019generative, rudolph2024trainability, kiwit2023application}. Here, we describe our proposal for constructing target distributions for these models, where the complexity is tunable in several ways.
This construction relies on MNs, where the graph structure can be defined by the user. In general, the ``difficulty'' of the learning problem is proportional to the clique sizes of the MN. The most general case is a complete graph with a single maximal clique that corresponds to explicitly encoding the probability to each global state.

Given an undirected graph $\mathcal{G}$, a set of cliques $\mathcal{C}_{\mathcal{H}}$, and a classical (pseudo)random number generator, we construct a target MN $\mathcal{H}$, as in Def.~\ref{def:mn}, through the following steps:
\begin{enumerate}
    \item To each clique $C \in \mathcal{C}_{\mathcal{H}}$, we assign a factor as a set of $ 2^{|C|}$ numbers obtained by querying the classical generator $ 2^{|C|}$ times.
    \item Next, we calculate the unnormalized measures of each global assignment by multiplying the corresponding element of each factor table.
    \item To get the probability of each assignment, we then normalize the measures by the partition function.
    \item Finally, we sample this joint probability distribution classically several times to construct the training dataset $\mathcal{D}$. 
\end{enumerate}

As the size of the graph increases, this procedure becomes highly inefficient, because of the exponential size of the space. In these cases, instead of calculating the target probabilities exactly, steps $2-4.$ can be replaced by the use of approximate sampling techniques to sample $\mathcal{H}$ directly~\cite{koller2009probabilistic}, e.g., by using Markov chain Monte Carlo methods. 

Similar steps can be taken for BNs as well, but there one chooses $2^{|Pa^{\mathcal{G}}_{X_i}|}$ random probabilities for each random variable $X_i$.

The main complexity of the target problem comes from the graph topology and the size of the cliques. This means that given a graph structure and its maximal clique factorization, we can increase the complexity of the corresponding target distribution by introducing additional edges and considering the maximal cliques of the new graph. 
The second complexity factor comes from the classical generator that assigns factor values to the cliques. 
Throughout our numerical investigations, we sample these values uniformly at random in an IID fashion from a positive range of real numbers. Alternatively, one can consider sampling the factor values from a more complex distribution, as long as classical sampling can be performed efficiently.

\subsection{Numerical Experiments}
\label{sec:numerics}

\begin{table*}[t]
\centering
\renewcommand{\arraystretch}{1.3} 
\setlength{\tabcolsep}{10pt}
\begin{tabular}{@{}lllll@{}}
\toprule
\textbf{Model} & \textbf{Classification} & \textbf{Circuit Topology} & \textbf{Maximal Gate Locality} \\ \midrule
\textbf{QCMRF} (Ours) & Problem-informed & Matches target MN (undirected) & Maximal clique size \\
\textbf{QCIBM} & Problem-agnostic & All-to-all connectivity (undirected) & $2$-local \\
\textbf{BBQC} & Problem-informed & Matches target BN (directed) & Maximal number of parents \\
\bottomrule
\end{tabular}
\caption{Comparison of the quantum generative models used in the numerical experiments, highlighting their classification, structural dependencies and gate locality.}
\label{tab:models}
\end{table*}

\begin{figure*}[t]
    \centering
    \includegraphics[width=.9\textwidth]{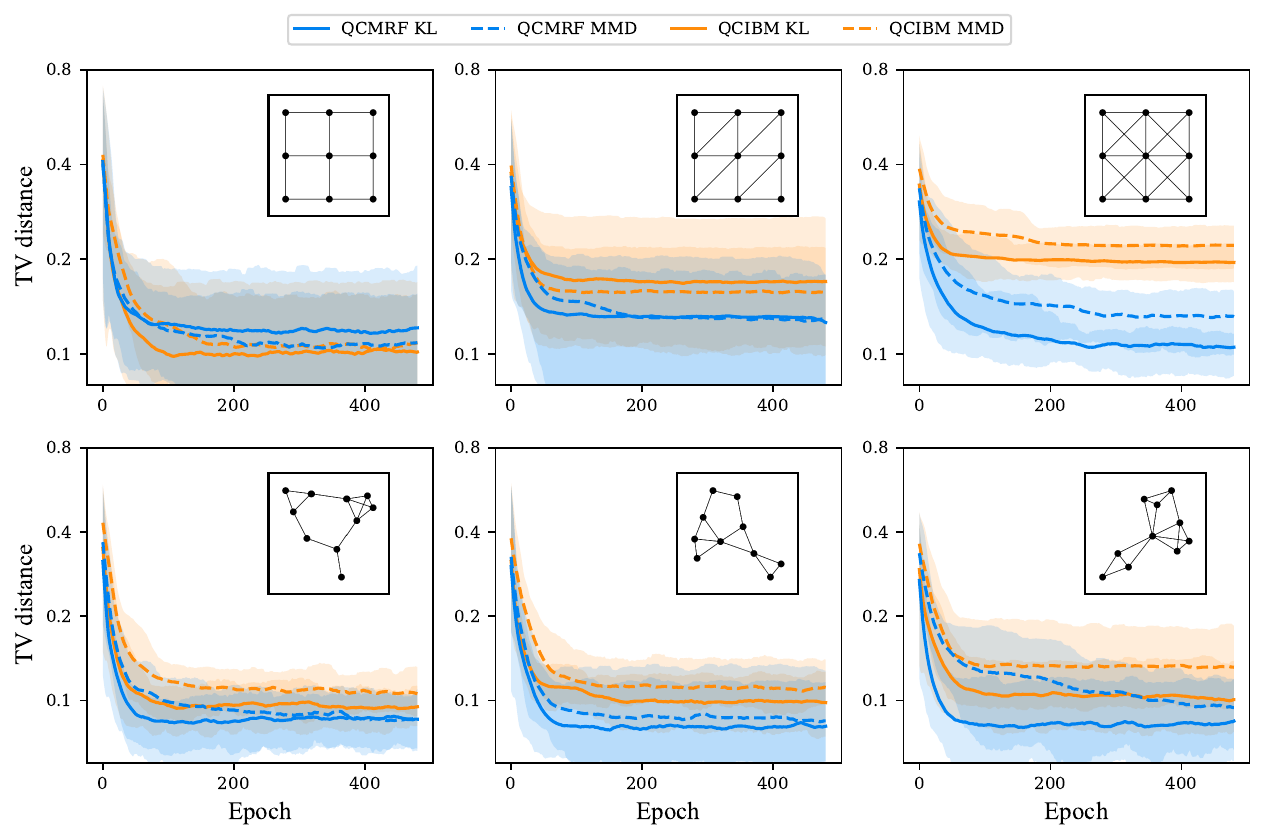}
    \caption{\textbf{QCMRF benchmark results against the problem-agnostic QCIBM using KL divergence and Maximum Mean Discrepancy (MMD) loss functions.} In the case of grid topologies (top row), as the complexity of the problem increases by using larger maximal cliques, the performance of the QCIBM decreases, while QCMRFs are either unaffected by this or even improve, as they also increase their complexity. The structured random graphs (bottom row) have densely connected communities, but the connectivity of these subgraphs is sparse. In this case, the QCMRFs still show good performance, significantly reducing the number of trainable parameters, but also improving the performance when the communities are connected by a node with large centrality measures (last graph).}
    \label{fig:double-exp}
\end{figure*}

Here we present two types of numerical experiments: the first kind aims to show that our QCMRF performs better than the problem-agnostic QCIBM on structured MNs; in the second set, we compare its performance to BBQCs. In the latter case, we consider loop graphs that first need to be triangulated in order to implement the corresponding BBQC models.
For ease of reference, Table~\ref{tab:models} summarizes the defining architectural features of these models.

In all experiments, trainings are carried out with two different cost functions: the KL divergence as in \eqref{eq:kldiv} and the MMD loss as in \eqref{eq:mmd}, where the kernel function is calculated as the average of $3$ different Gaussian kernels with $\sigma \in \{0.25, 10, 10^3\}$. The KL divergence has access to the exact target probability distribution, while the MMD loss can only access a finite training set.

The quantum circuit simulations are carried out with the Pennylane software package~\cite{bergholm2018pennylane}, and optimized with Adam~\cite{kingma2014adam}, with learning rate $0.1$. We train all models for $500$ epochs.

While we use only a finite number of shots for training, the TV distance between the model and target distributions is calculated analytically in each step.
We run multiple experiments with different random factor values and analyse the average performance of all models with both cost functions as measured in the exact TV distance. For better visualization, we also average over a window of $20$ training epochs.

\begin{figure*}[t]
    \centering
    \includegraphics[width=.9\textwidth]{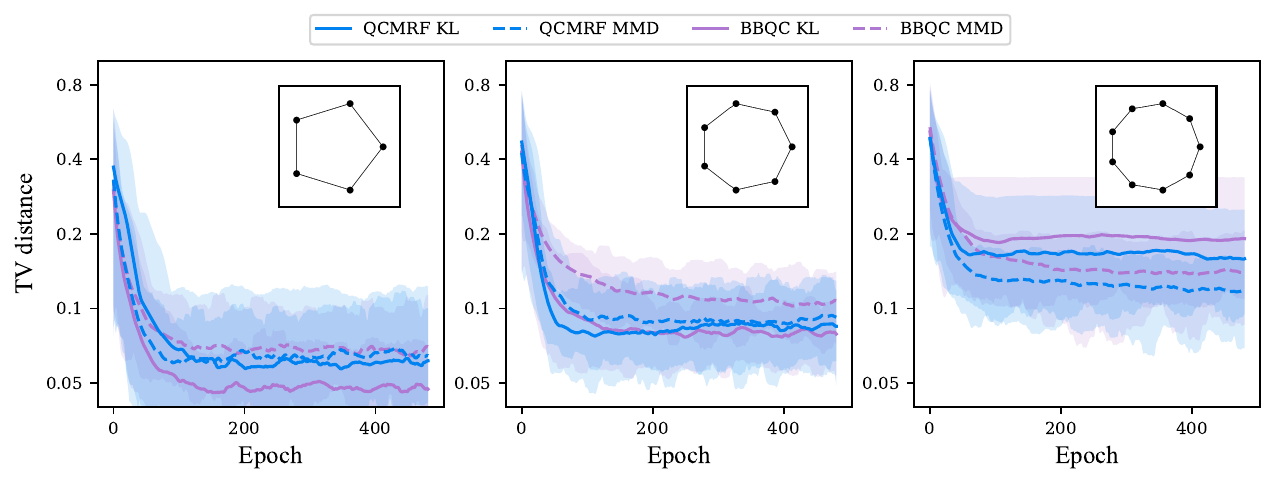}
    \caption{\textbf{Validating results against BBQC models on loop graphs.} The QCMRFs reach comparable performance to the BBQCs, while requiring less parameters and much shallower circuits. This difference is associated with the fact that these graphs have to be triangulated before implementing the corresponding BBQCs.}
    \label{fig:loop-exp}
\end{figure*}

\subsubsection{Benchmarks against QCIBMs}

To demonstrate the superiority of our model compared to the problem-agnostic QCIBM with all-to-all connectivity, first we present simulation results based on MNs with grid-like topology, always considering the maximal clique factorization. The number of training samples for the MMD loss along with the number of quantum circuit evaluations was set to $10^4$. We consider $5$ sets of uniformly random factor values and take the average performance over these.

All parameters of both models are initialized to $0$, as in this setting, the model starts the training in the equal superposition of all basis states. This strategy proved to be better than random initialization.

The results are shown in the top row of Fig.~\ref{fig:double-exp}. We start with a $3\times3$ grid, which defines a pairwise MN, meaning that the corresponding QCMRF incorporates only $2$-local interactions. Here, both models have similar performance, while our QCMRF reduces the number of trainable parameters from $72$ to $48$. 

We continue by introducing additional edges to the grid. All the maximal cliques of the second graph have size $3$, leading to a $3$-local Hamiltonian that can capture higher-order correlations. Here we can already see some separation in the performance, while still having less parameters ($60$) than the QCIBM. In the last case, the cliques are of size $4$, which increases the number of trainable parameters in the QCMRF circuit to $76$. In this case, our model significantly outperforms the $2$-local QCIBM. 

This series of experiments shows that as we increase the connectivity of the graph and the sizes of its maximal cliques, the distribution becomes harder to learn, as it is reflected in the performance of the problem-agnostic QCIBM. However, the performance of our problem-specific QCMRF is either unaffected by this change, or it performs even better, as its complexity also increases with the underlying MN. This also proves that using higher-order Hamiltonians can actually help in capturing higher-order correlations between the random variables of the MN.

Next, we focus on random graphs that are globally less structured, thus being closer to naturally occurring topologies. Here we explore the role of communities. By communities, we refer to dense subgraphs that are sparsely connected between each other. The results are shown in the bottom row of Fig.~\ref{fig:double-exp}, where the training is done similarly to the previous experiments. Here our QCMRFs reach better performance than the problem-agnostic QCIBMs, while significantly reducing the number of trainable parameters, since they exploit the sparsity of the graph. For these $10$-node graphs, the corresponding QCIBM has $85$ parameters, while the QCMRF circuits only have $60$, $58$ and $66$ respectively. Furthermore, in the third graph, where the communities are connected by a node with large centrality measures, the QCMRF significantly outperforms the problem-agnostic case. These experiments further prove the usefulness and viability of our Ansatz design approach.

We also showcase the scaling of the performance separation with the number of qubits for a triangle chain graph. These results based on exact simulation (infinite shot limit) are shown in Appendix~\ref{appendix:numerics}, where we also present a minimal numerical comparison to classical methods.

\subsubsection{Benchmarks against BBQCs}

Next, we validate our model with BBQCs on non-chordal MNs. In these cases, the undirected graph first has to be triangulated and turned into a directed acyclic graph. Knowing the structure of this BN, one can implement the corresponding BBQC that is able to capture the target distribution exactly. However, this process is very costly: the triangulation of the graph itself is a hard problem, and it can introduce a large number of edges that leads to the introduction of additional variables. Consequently, the corresponding BBQC can have many more trainable parameters and the circuit depth can be significantly larger. 

For this comparison, we considered loop graphs that are easy to triangulate. Since with $0$ parameter initialization, the starting probability distribution is different for QCMRF and BBQC circuits, here we started with random values to have a more fair comparison. The number of shots was set to $10^3$ for all experiments, along with the size of the training set for the MMD-based optimizations. We ran all experiments with $10$ sets of uniformly random factor values and visualized the average performance in Fig.~\ref{fig:loop-exp}.

All these loop-graphs can be used to implement QCMRF circuits, while to define the corresponding BBQC, the triangulation introduces $n-3$ new edges. Due to this fact, the number of trainable parameters along with the depth of the circuit is significantly increased. In all cases, our MN-based model reaches similar performance to BBQCs, while having much lower cost of implementation, which further highlights the usefulness of our model class. For a deeper comparison between the implementation costs of these models, we refer the reader to Appendix~\ref{appendix:cost-analysis}.

\subsection{Trainability}
While MNs are capable of representing any probability distribution, we expect that not all types of networks lead to efficiently trainable QML models. For this reason, we conduct a preliminary analysis of the trainability properties of QCMRFs. In particular, we study numerically the scaling of the MMD cost function variance with the number of qubits (or nodes). Having fixed the graph type, for each number of nodes, we define $10$ sets of uniformly random factor values and in each case, we evaluate the variance using $10^4$ sets of random circuit parameters. The cost value is calculated with a training set of $10^4$ samples and $10^4$ quantum circuit evaluations.

\begin{figure}[htbp]
    \centering
    \includegraphics[width=.4\textwidth]{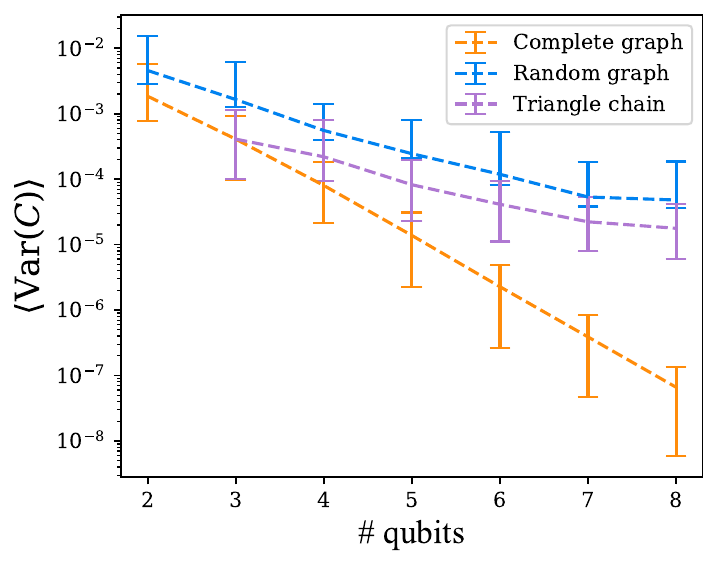}
    \caption{\textbf{Scaling of the cost variance for $3$ types of MN graphs.} Dashed lines connect the average variance over $10$ sets of random factor values, while the error bars show the minimum and the maximum for a given graph-type and number of qubits.}
    \label{fig:trainability}
\end{figure}

A complete graph having maximal clique factorization corresponds to no structure, since the probability distribution has $2^n-1$ degrees of freedom. As shown in Fig.~\ref{fig:trainability}, according to our numerics, the cost function variance vanishes exponentially in the number of qubits in this case, which also indicates the presence of deterministic barren plateaus~\cite{arrasmith2022equivalence}. To study sparser (but still dense) graphs, we concentrate first on Erd\H{o}s-R\'enyi graphs with $p=0.5$ edge probability, since in this case, the expected size of the largest clique is $\mathcal{O}(\log n)$ with high probability. Our simulations clearly show that the variance exhibits much better scaling in this case, than the conclusively exponential decay before. This is also true for triangle chain graphs, where we have $n-2$ cliques of size $3$ in total. While cost concentration and gradient concentration go hand in hand in variational quantum algorithms, polynomial scaling in the cost variance can be induced by a few parameters with good behaviour~\cite{arrasmith2022equivalence, miao2024equivalence}. Consequently, such a polynomial scaling does not show the empirical absence of barren plateaus, but it is a good first indicator of better trainability.

These investigations lead us to the definition of a subclass of MNs that restrict the class of problems that our problem-informed framework can be most useful for.

\begin{definition}[Efficient MN representation]\phantom{Anddd}
A probability distribution $P_{\Phi}$ over $n$ binary random variables is said to have an efficient MN representation, if $P_{\Phi}$ factorizes according to a Markov network $\mathcal{H}$ with
$$
\sum_{C\in\mathcal{C}_{\mathcal{H}}}2^{|C|} \in \mathrm{poly}(n),
$$
where $\mathcal{C}_{\mathcal{H}}$ is the set of cliques of $\mathcal{H}$.
\label{def:efficient_mn}
\end{definition}

It follows naturally from our definitions that QCMRF circuits, corresponding to efficient MN representations, have depth $\mathrm{poly}(n)$.

\subsection{Theoretical Foundations for Quantum Advantage}

Quantum advantage, in the context of generative learning, can have various flavours: it can show improvement in the precision (with respect to a distance metric); it can refer to faster convergence; or even an improvement in the number of training samples needed. In order to formulate all these cases in a single definition, we first need to define what we mean by a class of distributions being efficiently learnable:

\begin{definition}[$(d, \varepsilon, k, C)$-learnable]
For a metric $d$, $\epsilon > 0$, $k > 0$ and complexity class $C$, a class of distributions $P^n$ is called $(d, \varepsilon, k, C)$-learnable, if there exists an algorithm $\mathcal{A} \in C$ that, given $0 < \delta < 1$ as input and having access to a dataset $\mathcal{D}$ of size $k$ sampled from any $P \in P^n$, outputs with probability at least $1-\delta$ a representation of a distribution $P_{\mathcal{M}}$ satisfying $d(P, P_{\mathcal{M}})\leq \varepsilon$. $\mathcal{A}$ should run in time $\text{poly}(1/\varepsilon, 1/\delta, n)$.

\label{def:learnable}
\end{definition}

With this, we extended the definition from~\cite{coyle2020born} with the sample complexity, and now formulate quantum learning advantage:

\begin{definition}[Quantum Learning Advantage]
An algorithm $\mathcal{A} \in BQP$ is said to have quantum learning advantage, if there exists a class of distributions $P^n$ for which there exists $d$, $\epsilon$, $k$ such that $P^n$ is $(d,\varepsilon, k, BQP)$-learnable, but not $(d,\varepsilon, k, BPP)$-learnable, where this concept of learnability is defined in Def.~\ref{def:learnable}.

\label{def:qla}
\end{definition}

Besides learning advantage, generative QML models also have the potential to exceed classical methods for sampling the learned distribution:

\begin{definition}[Quantum Advantage in Sampling]
Given a probability distributions $P_{\mathcal{M}1}$ satisfying $d(P, P_{\mathcal{M}1})\leq \varepsilon$ (for some $d$ metric and $\varepsilon > 0$), a quantum algorithm $\mathcal{A}_1 \in BQP$ is said to have quantum advantage in sampling from the distribution $P$, if $\mathcal{A}_1$ can efficiently sample $P_{\mathcal{M}1}$ and no classical algorithm $\mathcal{A}_2 \in BPP$ can sample $P_{\mathcal{M}2}$ efficiently, where $d(P, P_{\mathcal{M}2})\leq \varepsilon$.

\label{def:samp-adv}
\end{definition}

In the following, we concentrate on this second definition and argue that, since our QCMRF class contains the class of QAOA and IQP circuits, it can also produce distributions that are thought to be classically hard. For this, we assume  that the joint distribution of a target MN is learnable by both an arbitrary classical model and a QCMRF to a given precision $\varepsilon$ and analyse the complexity of sampling the trained models.
Previous works~\cite{coyle2020born, zhu2022generative} presented similar arguments, relying on the results of Refs.~\cite{bremner2011classical, farhi2016quantum}.
We start by sketching current results regarding the hardness of sampling QAOA circuits, then based on this, we present our conjecture about the possible quantum advantage in sampling for our model.

The QAOA was proposed in Ref.~\cite{farhi2014quantum} for approximately solving combinatorial optimization problems on a quantum computer. 
 In Ref.~\cite{farhi2016quantum}, the authors proved  that efficient classical sampling of the output distribution of the $1$-level QAOA circuit implies the collapse of the polynomial hierarchy to the third level. While the argument was constructed with a $2$-local QAOA circuit, the proof stands for higher-order $Z$ interactions as well, all these being diagonal operators. On the other hand, this was shown for multiplicative error only, meaning  that if the target distribution (being the one defined by the QAOA circuit) $P_{QAOA}$ and the classical model distribution $P_{\mathcal{M}}$ satisfy the bound
\begin{equation}
    |P_{\mathcal{M}} (x) - P_{QAOA}(x)| \leq \varepsilon P_{QAOA}(x), 
\end{equation}
then the polynomial hierarchy collapses to its third level. In our framework, however, we mostly concentrated on the TVD, which is equivalent to additive error.
Multiplicative error is a stronger constraint, thus proving hardness with bounded additive error is a harder task. 
These results using multiplicative error were extended in Ref.~\cite{dalzell2020many} from worst case to average case hardness for $1$-level QAOA circuits. Finally, in Ref.~\cite{krovi2022average}, the author proved average case hardness with additive error, which, up to our knowledge, is the strongest (additive error) result in connection with the weak simulation of $1$-level QAOA circuits. 

Since the class of QCMRF circuits contains the $1$-level QAOA circuits, this larger class also contains classically hard cases.
To make this transition explicit, we note that fixing the parameters of the final single-qubit unitaries in a QCMRF circuit as $U_f(\boldsymbol{\Gamma}, \boldsymbol{0}, \boldsymbol{0})$ (where $\boldsymbol{0}$ denotes the all-$0$ vector), and $\Gamma_k = \gamma$ for all $k$, we arrive at a $1$-level QAOA circuit with a standard $X$-mixer, where the structure of the diagonal operator $U_Z$ is dictated by the target MN. Similarly, fixing the parameters $\Gamma_k = \Sigma_k = \pi/(2\sqrt{2})$ for all $k$ and $\boldsymbol{\Delta} = \boldsymbol{0}$, the QCMRF strictly reduces to an IQP circuit, which has also been rigorously proven to be classically intractable to sample from~\cite{bremner2016average}.
These results highlight the potential of QCMRFs to provide sampling advantage in the inference stage even for sparse graphs, such as $2D$ grids, leading to the following conjecture.

\begin{conjecture*}
The class of QCMRF circuits  that can learn probability distributions with an efficient MN representation (as in Def.~\ref{def:efficient_mn}), also contains classically hard cases, yielding a quantum advantage in sampling.
\end{conjecture*}

\subsection{Scalability demonstration}

\begin{figure*}[t]
    \centering
    \includegraphics[width=.9\textwidth]{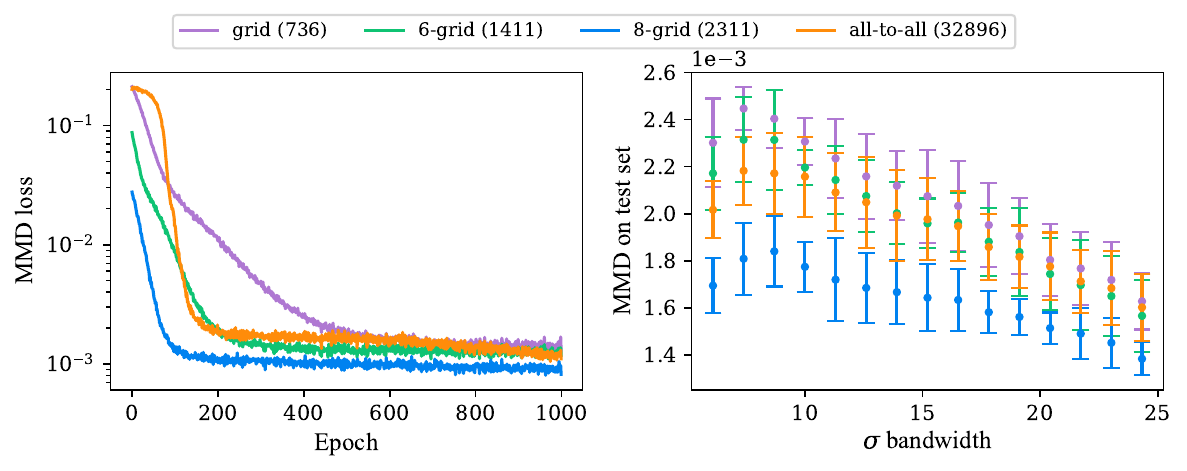}
    \caption{\textbf{Scalability study using IQP circuits.} We trained $3$ IQP models with the QCMRF ansatz and different MN assumptions and their problem-agnostic counterpart (all-to-all) for comparison. Besides the training curves (left) the test results of the trained models are also shown (right). The test results were obtained by estimating the MMD of the trained models with respect to a test set, and a varying $\sigma$ bandwidth.}
    \label{fig:iqp}
\end{figure*}

Finally, to demonstrate the scalability of the QCMRF Ansatz design approach, we consider a more restricted version of our model class, by replacing the final layers of trainable $U_f$ gates with Hadamard gates. This simplification effectively transforms our model to a form of IQP-QCBM, as explained in the previous section. It was recently shown that these models can be efficiently trained on a classical computer at scale, but can still require a quantum device for efficient sampling in the inference stage~\cite{recio2025train}. The classical training technique relies on the reformulation of the MMD loss function using local expectation values that can be estimated classically. In this experiment, we show that our method for constructing the $U_Z$ operator is beneficial in IQP circuits as well.

We consider the USPS dataset of $16\times16$ grayscale images of handwritten digits~\cite{hull1994database}. These images are first flattened and binarized to construct the training and test datasets composed of $19999$ and $1500$ samples of $256$-long bitstrings, respectively. Following the approach from Ref.~\cite{recio2025train}, we can define the $U_Z$ operator using all $1$- and $2$-local gates, which corresponds to the QCIBM Ansatz (here we refer to this Ansatz as all-to-all). However, it is a well known fact that in image data, pixels that are close together tend to be highly correlated, leading to the use of MNs with grid topologies~\cite{besag1986statistical}. Therefore, we construct the problem-informed models using the three grid structures from Sec.~\ref{sec:numerics}: a simple pairwise grid, a $6$-grid with cliques of size $3$, and an $8$-grid with cliques of $4$. These MN assumptions lead to heuristic Ansatz constructions influenced by basic domain knowledge. Furthermore, these grid-structures inherently define their own maximal cliques by design, therefore, no costly clique-finding process is needed.

We train the IQP models corresponding to all $4$ Ans\"atze using the MMD loss function with $3$ bandwidths $\sigma \in \{6.16, 12.33, 24.66\}$. The learning rate was set to $0.001$ and we trained for $1000$ epochs. Finally, we tested the trained models by evaluating the MMD with respect to the test set using different $\sigma$ bandwidths and averaging over $5$ independent estimates. All trainings and tests were performed using the IQPopt package~\cite{recio2025iqpopt}. The final results are shown in Fig.~\ref{fig:iqp}, where we also show the number of trainable parameters for each model. These results demonstrate that using our simple MN assumptions, we did not only reduce the number of trainable parameters significantly, but in the case of the $8$-grid Ansatz, this also sped up convergence and reached better test results than the problem-agnostic all-to-all Ansatz.

In order to isolate the effect of the problem-informed architectural inductive bias, we evaluated all models under identical hyperparameter settings and training budgets. We note that the all-to-all Ansatz is not necessarily bottlenecked by a lack of theoretical expressivity; with extensive hyperparameter tuning -- specifically, by employing a higher learning rate and extending the training duration -- this problem-agnostic Ansatz can eventually achieve comparable final performance to the problem-informed model with $8$-grid topology. However, the fact that we achieve this performance under standard settings with a fraction of the parameters rigorously validates our core thesis: embedding the MN structure into the circuit topology fundamentally resolves the severe trainability bottlenecks associated with problem-agnostic generative models.

\section{Conclusion \& Outlook}
\label{sec:conclusion}

In this work, we highlighted the potential of probabilistic graphical models for generative QML. We introduced a framework for constructing quantum circuit Born machine Ans\"atze  that respect the structure of the Markov network describing the underlying problem. A novel problem construction process was presented for benchmarking generative QML models, where the complexity of the learning task can be tuned in various ways. This benchmarking framework is capable of constructing explicit distribution learning problems and more realistic tasks based on limited samples.

Our numerical experiments demonstrated that our model, called quantum circuit Markov random field, is capable of capturing higher-order correlations between the binary random variables of the corresponding MN. 
This can significantly improve the performance in the case of higher-order target models, while potentially reducing the number of trainable parameters on sparse graphs, compared to problem-agnostic approaches.
We show in Appendix~\ref{appendix:cost-analysis} that in practical cases, such as those involving grid topologies, this asymptotically reduces the overall implementation cost, as measured in both the number of $R_Z$ rotations and $CNOT$ gates.
We further validated our model with basis-enhanced Bayesian quantum circuits on non-chordal MNs, since these BN-based models are able to express the target distribution exactly. 
The QCMRFs reached the performance of BBQCs on small loop-graphs with fewer parameters and significantly shallower circuits. All these experiments were conducted using the KL divergence and MMD loss functions, to demonstrate both exact distribution learning tasks and more practical generative learning based on limited training samples.

A preliminary numerical analysis of trainability was presented  that introduced an important constraint on the sparsity of the potential MNs of interest.
We formulated two definitions of quantum advantage, relevant in the context of generative models, where the first concentrated on a learning advantage (Def.~\ref{def:qla}), and the second focused on efficiently sampling from the learned distribution (Def.~\ref{def:samp-adv}). 
We presented an argument in the second setting, highlighting the potential of our model to offer improvements over classical methods, since it contains the class of QAOA circuits that are known to be hard to sample classically.
We also believe that this connection between classical MNs and QAOA-type circuits opens up an interesting direction for further investigations. 
Finally, we demonstrated the applicability and scalability of our model design framework in a practical setting using $256$-qubit IQP circuits on a problem, where the exact MN structure is not readily available, but the structural assumptions are made using basic domain knowledge. It is important to note that for arbitrary, unstructured graphs, maximal clique enumeration is an NP-hard problem, meaning this approach is best suited for domains where the graph structure is constrained or defined by known spatial/relational priors.

The proposed quantum circuit model is not hardware-efficient, however, since the core of the Ansatz is composed of commuting gates, it can be further optimized to fit hardware connectivity and other constraints. As shown in Appendix~\ref{appendix:cost-analysis}, simple grid structures have asymptotically efficient resource requirements, and these circuits can be implemented with minimal overhead, e.g., on superconducting devices with lattice topologies. Furthermore, in early-fault tolerant settings, where limited error correction is available, such diagonal circuits provide good candidates for early quantum advantage~\cite{hangleiter2025fault}, as they can be implemented more efficiently than generic quantum algorithmic primitives.

While we concentrated on learning a target distribution to high accuracy, this alone is not enough to characterize the performance of generative models. Another important factor is the model's ability to generalize, rather than memorize the training data~\cite{gili2023quantum}. In the context of PGMs, this can be investigated with a training dataset of limited size, assessing the trained model's ability to generate valid but unseen samples.
We also remark  that while our QCMRF shows significant improvement compared to problem-agnostic QCIBMs on several small examples, BBQCs show better performance on chordal graphs, where both PGM-based models have the same number of trainable parameters. This means  that for chordal graphs, BN-based models are better at capturing the target distribution, although requiring much deeper circuits.
Finally, what can be seen as a setback of our framework is that it assumes knowledge of the graph structure along with its cliques. While this seems like a major obstacle, this knowledge is readily available in various use cases, and educated assumptions can be made by domain experts. Furthermore, this generative QML approach can be used together or as a subroutine of classical structure learning algorithms~\cite{schluter2014survey}.

Possible extensions of our model could include replacing the final set of general one-qubit unitaries $U_f(\Gamma, \Delta, \Sigma)$ with one-parameter rotations (e.g., $R_X$ rotations), as well as using multiple layers to have higher resemblance to QAOA circuits and to reach the overparametrized regime. 
Finally, we also remark that, while our model design framework was constructed in the context of QCBMs, this idea can also help in designing other generative QML models, e.g., using the QCMRF as the generator circuit in QGANs, as demonstrated in Appendix~\ref{app:gan}.

\section*{Acknowledgements}

The authors would like to thank the support of the Hungarian National Research, Development and Innovation Office (NKFIH) through the KDP-2021 and KDP-2023 funding scheme, the Quantum Information National Laboratory of Hungary and the grants TKP-2021-NVA-04, TKP-2021-NVA-29 and FK 135220. Z.Z. was partially supported by the Horizon Europe programme HORIZON-CL4-2022-QUANTUM-01-SGA via the project 101113946 OpenSuperQPlus100 and the QuantERA II project HQCC-101017733.
The authors also acknowledge the computational resources provided by the Wigner Scientific Computational Laboratory (WSCLAB). \\[-3mm]

\appendix

\section{Ansatz design}
\label{appendix:qcmrf-ansatz}
Here we give a more detailed description of our QCMRF Ansatz design process.
Let us consider a single clique $C$, with factor $\phi_C$, and embed the corresponding marginal probability distribution in the diagonal of a $2^{|C|}\times 2^{|C|}$ matrix as
\begin{equation}
    \begin{aligned}
    H_C & = \mathrm{diag}(P_{\phi_C}) \\
    & = \frac{1}{Z_{\phi_C}} \sum_{x<2^{|C|}} \phi_C(x) \ketbra{x}{x} \\
    & = \frac{1}{Z_{\phi_C}} \sum_{x<2^{|C|}} \phi_C(x) \bigotimes_{i}\frac{1}{2}(\mathds{1}_i + (-1)^{x_i}Z_i).
    \end{aligned}
\end{equation}
Here $Z_{\phi_C}$ denotes the normalizing constant of the marginal distribution, $|C|$ denotes the size of clique $C$, and in the last line, we used the Pauli expansion of the projectors.

We can reparametrize this linear combination of Pauli strings as 
\begin{equation}
\begin{split}
H_C(\boldsymbol \alpha) & = \alpha_0 \bigotimes_i \mathds{1}_i + \alpha_1 Z_0 + \alpha_2 Z_1 + \dots \\
& + \alpha_{2^{|C|}-1} \bigotimes_i Z_i.
\end{split}
\end{equation}
After discarding the identity term, we consider the unitary $U_C(\boldsymbol \alpha) = \exp(-iH_C(\boldsymbol \alpha))$. At this point, it is easy to see  that if this operator acts on the initial state as $U_C(\boldsymbol \alpha) \ket{+}^{\otimes n}$, then the marginal probability distribution is encoded into the phases of each computational basis state. We can also see  that there are $2^{|C|}-1$ parameters, which is exactly how many degrees of freedom a probability distribution over $|C|$ binary random variables has.

Going back to the full probability distribution $P_{\Phi}$  that factorises according to a MN as in Def.~\ref{def:mn}, we can embed this into the diagonal of a $2^n \times 2^n$ matrix of the form
\begin{widetext}
\begin{equation}
    \begin{aligned}
    H = \mathrm{diag}(P_{\Phi}) & = \frac{1}{Z_{\Phi}} \prod_{C\in \mathcal{C}} \left[\sum_{x<2^{|C|}}\phi_C(x)\bigotimes_{i\in C}\frac{1}{2}(\mathds{1}_i + (-1)^{x_{C(i)}}Z_i) \bigotimes_{i \notin C} \mathds{1}_i \right] \\
    & = \frac{1}{Z_{\Phi}} \sum_{x<2^{n}} \left[\prod_{C\in \mathcal{C}}\phi_C(x_C)\bigotimes_{i\in C}\frac{1}{2}(\mathds{1}_i + (-1)^{x_i}Z_i) \bigotimes_{i \notin C} \mathds{1}_i \right].
\end{aligned}
\end{equation}
\end{widetext}
The first two lines show equivalent definitions, where $C(i)$ denotes the index of node $i$ in clique $C$ and $x_C$ denotes the bits in the binary form of $x$  that correspond to clique $C$.

According to the previous argument, after reparametrizing this Hamiltonian as before, the unitary $U_Z(\boldsymbol \alpha) = \exp (-iH(\boldsymbol \alpha))$ is expressive enough to encode the marginal probability distributions over each clique. However, since the factors are not normalized in general, the joint distribution is not necessarily the same as the factor product of the marginal ones. For this reason, instead of extracting the phases of the state $U \ket{+}^{\otimes n}$, we perform a measurement on each qubit in any local basis, i.e., we apply general one-qubit gates before measurement, where these gates have independent trainable parameters. We expect  that this expressivity enhancement provides enough flexibility to compensate for the mismatch coming from the different scale of the marginal distributions.

We can also take the opposite route and analyse the output probability distribution of this quantum circuit model. We want to analyze the structure of the MN  that factorizes this probability distribution.
First, we can write the marginal probability of each random variable in the output distribution as 
\begin{equation}
    P(X_i) = - \frac{\langle Z_i \rangle - 1}{2},
    \label{eq:marg_prob}
\end{equation}
where $\langle Z_i \rangle$ denotes the expectation value of the Pauli-Z operator on qubit $i$. 
Thus, it suffices to examine the backward light-cone of observable $Z_i$ for all $i$. Due to the structure of the QCMRF, the entangling gates in the $U_Z$ operator can be ordered in such a way  that the light-cone of the local observable $Z_i$ only contains the qubits, it is directly connected to, through multi-qubit gates. This means  that the probability in Eq.\eqref{eq:marg_prob} can be written as a product of probabilities and conditional probabilities involving only the neighbours of node $i$. This is equivalent to the Markov property, stating that random variable $X_i$ is conditionally independent of all other variables given its neighbours. 

Since in $U_Z$, all operator terms commute, each ordering of the gates (generated by the tensor product of Pauli-Z operators) produces the same quantum state leading to the same output probability distribution. Therefore, the Markov property applies to all output random variables simultaneously. Due to the Hammersley-Clifford theorem for strictly-positive probability distributions~\cite{clifford1971markov}, the presence of this Markov property also means  that the joint probability distribution factorizes according to the corresponding MN.

\section{Implementation details}

\begin{figure*}[t]
\centering
\subfloat[ ]{
\label{subfig:UZdecomp}
\includegraphics[width=0.44\textwidth]{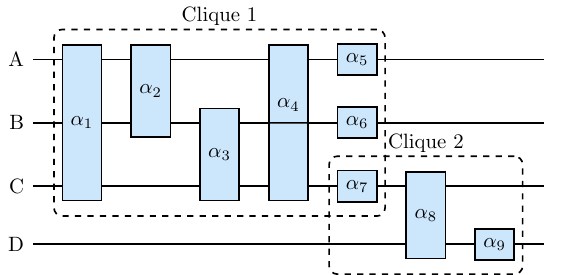}
}
\subfloat[ ]{
\label{subfig:zz-decomp}
\includegraphics[width=0.25\textwidth]{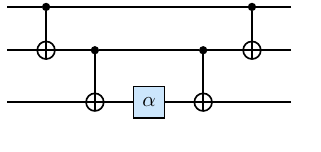}
}

\subfloat[ ]{
\label{subfig:clique-decomp}
\includegraphics[width=0.7\textwidth]{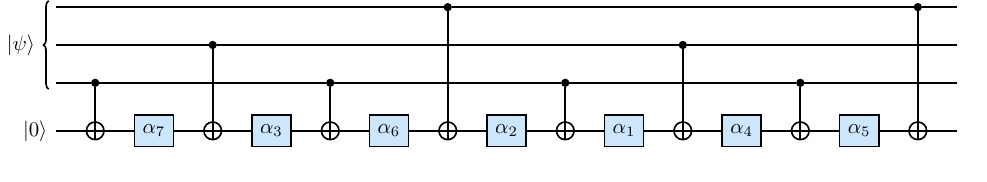}
}
  \caption{\textbf{Decomposition of $U_Z$ operators generated by many-body Ising Hamiltonians.} (a) Complete $U_Z(\boldsymbol \alpha)$ operator corresponding to MN in Fig.~\ref{subfig:mrf}. (b) Decomposition of $\exp{(-i \alpha ZZZ)}$, a $3$-body $MultiRZ$ gate parametrized by $\alpha$. (c) Implementation of all terms of a clique of size $3$ with an additional ancillary qubit. In all the figures, parametrized gates represent $k$-local $MultiRZ$ operators.}
\end{figure*}

In this section, we describe the implementation details of the PGM-based quantum models, including possible decompositions of multi-qubit operators into single-qubit and $CNOT$ gates.
\subsection{Quantum Circuit Markov Random Fields}
\label{appendix:qcmrf-implem}

Markov networks define higher-order Ising Hamiltonians as described in \ref{sec:qcmrf}, which generate QCMRF circuits composed of $MultiRZ$ gates. For the MN in Fig.~\ref{subfig:mrf}, the corresponding parametrized circuit is shown in Fig.~\ref{subfig:UZdecomp}. A $k$-local $MultiRZ$ gate can be implemented in linear depth with $2k$ $CNOT$ gates and a single-qubit $R_Z$ rotation. An example for this decomposition in shown in Fig.~\ref{subfig:zz-decomp}. There are several other alternative strategies for implementing these circuits, as explained in the context of QAOA in Refs.~\cite{glos2022space, bako2025prog}. For example, one could implement the circuit corresponding to a clique with $m$ random variables with significantly fewer $CNOT$ gates using an ancillary qubit, as shown in Fig.~\ref{subfig:clique-decomp}. This approach reduces the number of gates and the depth, while adding an ancillary qubit for each clique of size $>2$.

The implementation of the QCIBM circuit can be done similarly, only using $2$-local $MultiRZ$ gates  that do not need any ancillary qubits. 

\subsection{Bayesian Quantum Circuits}
\label{appendix:bbqc-implem}

\begin{figure*}[t]
\centering
\subfloat[ ]{
\label{subfig:bbqc}
\includegraphics[width=0.5\textwidth]{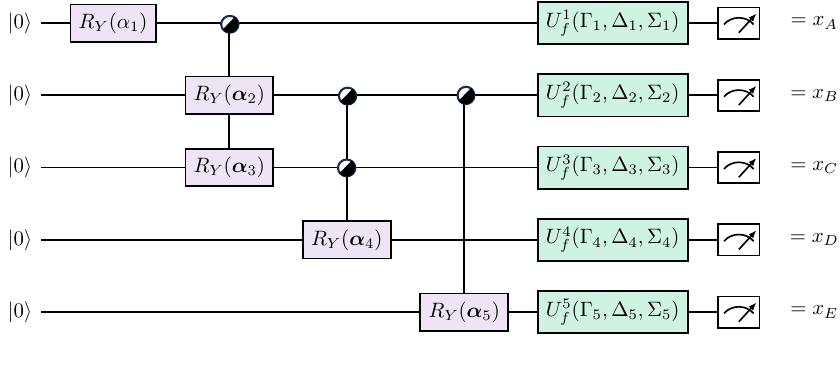}
}

\subfloat[ ]{
\label{subfig:ucry}
\includegraphics[width=0.6\textwidth]{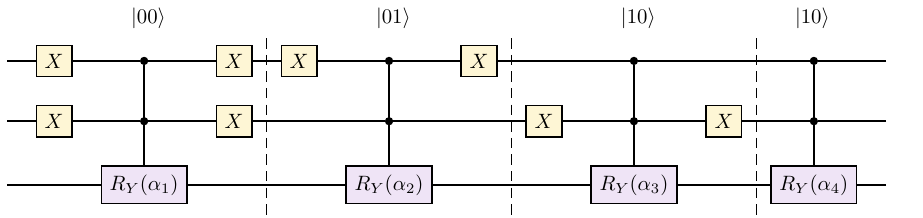}
}
  \caption{\textbf{Implementation of BBQC circuits.} (a) Complete BBQC circuit corresponding to the BN in Fig.~\ref{subfig:bayes} (b) Decomposition of a uniformly controlled $R_Y$ rotation gate, parametrized with parameter vector $\mathbf{\alpha}$, having two control qubits. The number of multi-controlled $R_Y$ rotations scales exponentially with the number of controls. Half-empty controls denote uniformly controlled gates.}
\end{figure*}

Here we start by describing the general idea of BQCs introduced in Ref.~\cite{low2014quantum}. This model associates a qubit to each binary random variable of the BN and then applies unitary operations in the following manner. First, parametrized single-qubit $R_Y$ rotation gates are applied to qubits for which the corresponding nodes have no parents. Then uniformly controlled $R_Y$ operations are performed on each qubit, where the control qubits correspond to the parents of the given node. Note  that since a node can have multiple parents, this can lead to gates with a large number of controls. The order of application of these unitaries has to follow two rules: every qubit can only be targeted once and after a qubit was used as a control, it cannot be targeted anymore. These rules ensure the compliance with the directed acyclic nature of the graph. The basis-enhancement introduced in Ref.~\cite{gao2022enhancing} applies further parametrized general $U_f(\Gamma, \Delta, \Sigma)$ gates to all qubits. The BBQC corresponding to the BN presented in \ref{subfig:bayes} is shown in Fig.~\ref{subfig:bbqc}.

While uniformly controlled gates provide an easily interpretable mapping from the BN graph to a parametrized quantum circuit, they cannot be implemented directly. The number of parameters for such a gate is $2^{n_c}$, where $n_c$ is the number of control qubits. In Refs.~\cite{borujeni2021quantum,PhysRevA.71.052330}, the authors gave several strategies to decompose these operators. For the purpose of this work, it is easiest to think of these as the decomposition in terms of (multi-controlled) $R_Y$ rotations and $X$ gates. This decomposition of a uniformly controlled $R_Y$ rotation with $2$ controls is shown in Fig.~\ref{subfig:ucry}. The multi-controlled gates can be further decomposed into single qubit rotation and $CNOT$ gates with additional ancillary qubits as described in Ref.~\cite{borujeni2021quantum} or by adapting the ancilla-free strategy of~\cite{da2022linear}.

\section{Resource estimation}
\label{appendix:cost-analysis}
Here we review the general cost of implementation, considering the number of parameters, qubits and the circuit depth. We compare these metrics for all three models, where possible.
To make resource scaling comparisons more explicit, in Table~\ref{tab:resource_examples}, we compare relevant resources of QCIBM circuits and QCMRFs with the grid topologies used in the numerical simulations. In the estimate of the number of $CNOT$ gates, we assume the ancilla-free implementation of Fig.~\ref{subfig:zz-decomp}. These examples also show that QCMRF circuits can be asymptotically more efficient than their problem-agnostic counterpart in practical scenarios. The noise robustness of these circuits depends on the underlying hardware architecture: in near-term devices implementing the models using physical gates, the leading factor is the number of $CNOT$ gates, while in early fault-tolerant devices with partial error-correction, the number of $R_Z$ rotations might be more relevant.

\begin{table}[h]
\centering
\renewcommand{\arraystretch}{1.3} 
\begin{tabular}{l|c|c|c}
\toprule
QCMRF & parameters & $CNOT$ & $R_Z$ \\ 
\midrule
grid & $6n - 2\sqrt{n}$ & $4n - 4\sqrt{n}$ & $3n - 2\sqrt{n}$ \\
6-grid & $9n - 8\sqrt{n} + 3$ & $10n - 16\sqrt{n} + 6$ & $6n - 8\sqrt{n} + 3$ \\
8-grid & $13n - 16\sqrt{n} + 7$ & \begin{tabular}[c]{@{}c@{}} $30n - 52\sqrt{n}$ \\ $+26$ \end{tabular} & $10n - 16\sqrt{n} + 7$ \\ \midrule
QCIBM & $n(n-1)/2 + 4n$ & $n(n-1)$ & $n(n-1)/2 + n$ \\ \bottomrule
\end{tabular}
\caption{Table comparing resource scaling for QCIBM and QCMRF with different grid topologies over $n$ qubits.}
\label{tab:resource_examples}
\end{table}
\subsection{Number of parameters}

An important factor in QML Ansatz design is the number of trainable parameters. This consideration leads to a delicate balance. On one hand, we want our model to have enough expressivity to capture the target distribution. On the other hand, we want to limit the number of parameters, to provide faster training, minimize noise and potentially escape barren plateaus~\cite{ragone2024lie, fontana2024characterizing}. 

In the case of the QCIBM and QCMRF, the number of parameters depends mostly on the number of terms in the Hamiltonian that generates the time evolution. However, these models cannot be compared directly, since in QCIBMs this number only depends on the number of qubits $n$ and in QCMRF circuits there is an explicit dependence on the graph topology and the clique factorization, but also the overlap of the cliques.

To represent a classical MN as complete factor-tables, one needs $\sum_{C \in \mathcal{C}} 2^{|C|}$ parameters, $|C|$ referring to the size of clique $C$.
Consequently, our Hamiltonian before reparametrization has exactly this many terms. This means that the parameter count in a QCMRF Ansatz is
\begin{equation}
    k_{QCMRF} \leq \sum\limits_{C \in \mathcal{C}} \sum\limits_{k=1}^{|C|} \binom{|C|}{k} + 3n = \sum\limits_{C \in \mathcal{C}} (2^{|C|} - 1) + 3n.
\end{equation}
We can see  that this number does not depend on the problem size, i.e. the number of binary random variables $n$ directly, only on the sizes of the cliques $l_C$. This means  that assuming the clique sizes of a given graph topology is constant, without explicit dependence on the number of nodes, the parameter count is $\mathcal{O}(n)$.

In the case of the problem-agnostic $2$-local QCIBM Ansatz, this number is $k_{QCIBM} = n (n-1) / 2 + 4n$, which is $\mathcal{O}(n^2)$.

It is worth mentioning that for pairwise MNs, meaning those defined by only pairwise interactions, lead to two-body terms between the qubits representing connected nodes and one-body Pauli-Z terms on all qubits. These give rise to QCMRFs with less or equal number of parameters than the corresponding QCIBM Ansatz, where we have equality only for complete graphs.

A similar analysis can be performed for BBQCs as well. Here the number of parameters is exactly
\begin{equation}
    k_{BBQC} = \sum\limits_{i=1}^n 2^{|Pa^{\mathcal{G}}_{X_i}|} + 3n,
\end{equation}
where $|Pa^{\mathcal{G}}_{X_i}|$ refers to the number of parents of the $i$-th node. 

Since the conversion between BNs and MNs can introduce additional edges, this conversion can also introduce additional parameters. This means that usually, the number of parameters in BBQCs is larger than in QCMRFs, when the underlying problem is defined by a MN. However, for chordal MNs, having maximal clique factorization, the number of parameters in the two model classes is equal.

\begin{figure*}[t]
    \centering
    \includegraphics[width=.9\textwidth]{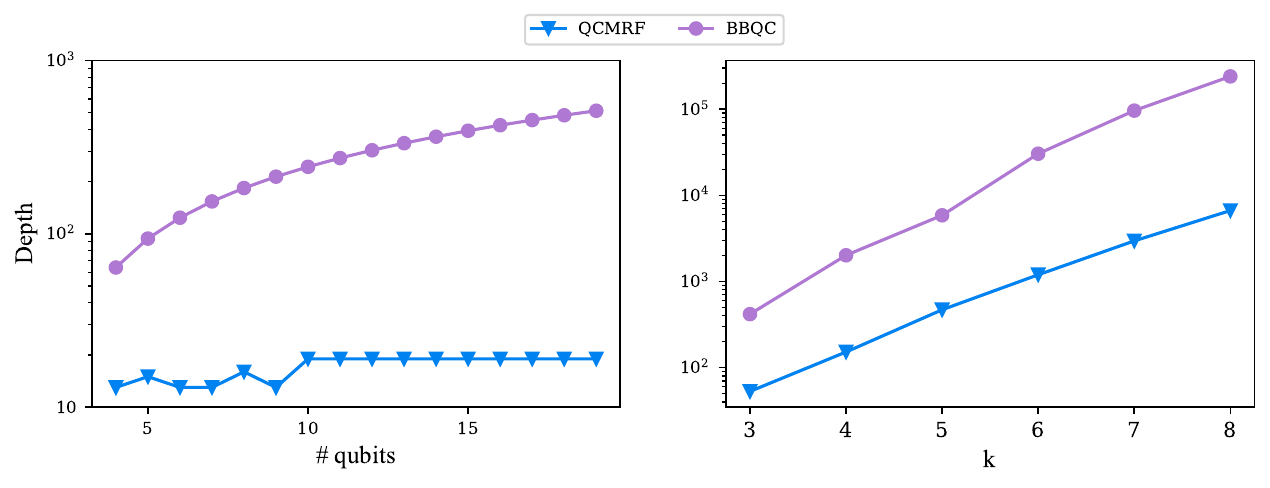}
    \caption{\textbf{Numerical comparison of circuit depth between QCMRF and BBQC models.} The quantum models corresponding to the given graph were transpiled using Qiskit with optimization level $3$. Scaling in the number of qubits for loop graphs (left). Scaling in the size of the maximal cliques for $k$-gram models (right).}
    \label{fig:depth}
\end{figure*}

\subsection{Number of qubits}
Since in all the models we consider, we map $n$ binary random variables to $n$ qubits, the number of qubits in all these cases is initially $n$, though certain gate decompositions require ancillary qubits.

For the QCIBM, since they only contain one- and two-qubit gates, we need no additional ancillas and the final number of qubits is $n$. In the case of QCMRF circuits we have at least two options, we can use no ancillas leading to the same number, or we can use $n_C$ ancillas, where $n_C$ refers to the number of cliques that are larger than $2$. In this latter case, we end up with $n + n_C$ qubits. We have even more freedom for BBQC models. We can use the ancilla-free implementation of multi-controlled gates from~\cite{da2022linear}, or we can use $l-1$ ancillary qubits to implement each multi-qubit gate with $l$ controls, where the ancillas can also be reused. It is also possible to interpolate between the two approaches, $n$ and $n+\sum_{i=1; |Pa^{\mathcal{G}}_{X_i}| \neq 0}^n (|Pa^{\mathcal{G}}_{X_i}| - 1)$. This lead to a trade-off between the number of qubits and circuit depth.

\subsection{Circuit depth}

Based on the implementation of multi-qubit gates, the depth of the circuit can also vary. In general, the more ancillas we use for decomposing these unitaries, the shallower the final circuit can be. For a fair comparison, here we only consider ancilla-free implementations and analyse the depth accordingly. We also assume full connectivity of qubits and parallel execution of gates that act on disjoint sets of qubits, where possible, and use parametrized one-qubit $U_f$ and $CNOT$ gates as our basis gate set.

QCIBM circuits being problem-agnostic, their depth can be estimated knowing only the number of qubits. The initial Hadamard gates along with the single-qubit $R_Z$ rotations and final $U_f$ gates can be implemented in constant depth. Since we assume parallel execution of gates and each two-qubit $\exp(-i\alpha ZZ)$ gate can be implemented in depth $3$, the depth of the whole circuit scales $\mathcal{O}(n)$.

For QCMRF circuits, the scaling is more complicated, since it depends on the clique-structure of the MN. Here we give an upper-bound on the depth, considering the Hamiltonian before reparametrization. For each clique $C$, we have $2^{|C|} - 1$ $MultiRZ$ gates, and there are ${|C| \choose k}$ $k$-local gates for each $k\leq |C|$. To minimize the depth, we can implement two gates in parallel in each step, such that the qubits they act on are complements of each other in the clique. In the ancilla-free case, each $k$-local gate can be implemented in $2k-1$ depth. This means that the depth required to implement such a clique is $\mathcal{O}(2^{|C|})$. If we assume $|\mathcal{C}|$ cliques in the graph and the size of the largest maximal clique is $l_m$, then the depth required is $\mathcal{O}(|\mathcal{C}| 2^{l_m})$. However, this is a crude upper bound in the worst case, not taking into account the parallel execution of gates of disjoint cliques. In practice, this scaling can be much better.

The analysis is similar for the BBQC models. Here for each node $X_i$ the number of multi-controlled $R_Y$ gates scales $\mathcal{O}(2^{|Pa^{\mathcal{G}}_{X_i}|})$. Another difference is that we cannot implement these unitaries in parallel, which further increases the depth. By fixing the largest number of parents as $n^p_m$, the depth scales $\mathcal{O}(n 2^{n^p_m})$ in the worst case. It is easy to see that for BNs and MNs based on chordal graphs, the size of the largest clique is on the order of the maximal number of parents. 

We compare these two models on the simple, yet illustrative example of a graph composed of a single triangle. This is obviously a chordal graph, and the number of trainable parameters is equal in both circuit. In this case, the depth of the QCMRF circuit is $16$. In the corresponding BN, the first node has no parents leading to depth $1$. The second node has $1$ parent, leading to $2$ single-controlled $R_Y$ gates, each implemented in depth $4$, thus depth $10$ with the $X$ gates. The final node has $2$ parents, meaning $4$ multi-controlled gates with two controls, each having depth $14$, thus depth $60$ in total. The final depth of this circuit is $72$ with the $U_f$ gates before measurement. This example shows that BBQCs can be much deeper than QCMRF circuits on chordal graphs, while this is not reflected in the number of trainable parameters.

\begin{figure*}[ht]
    \centering
    \includegraphics[width=.9\textwidth]{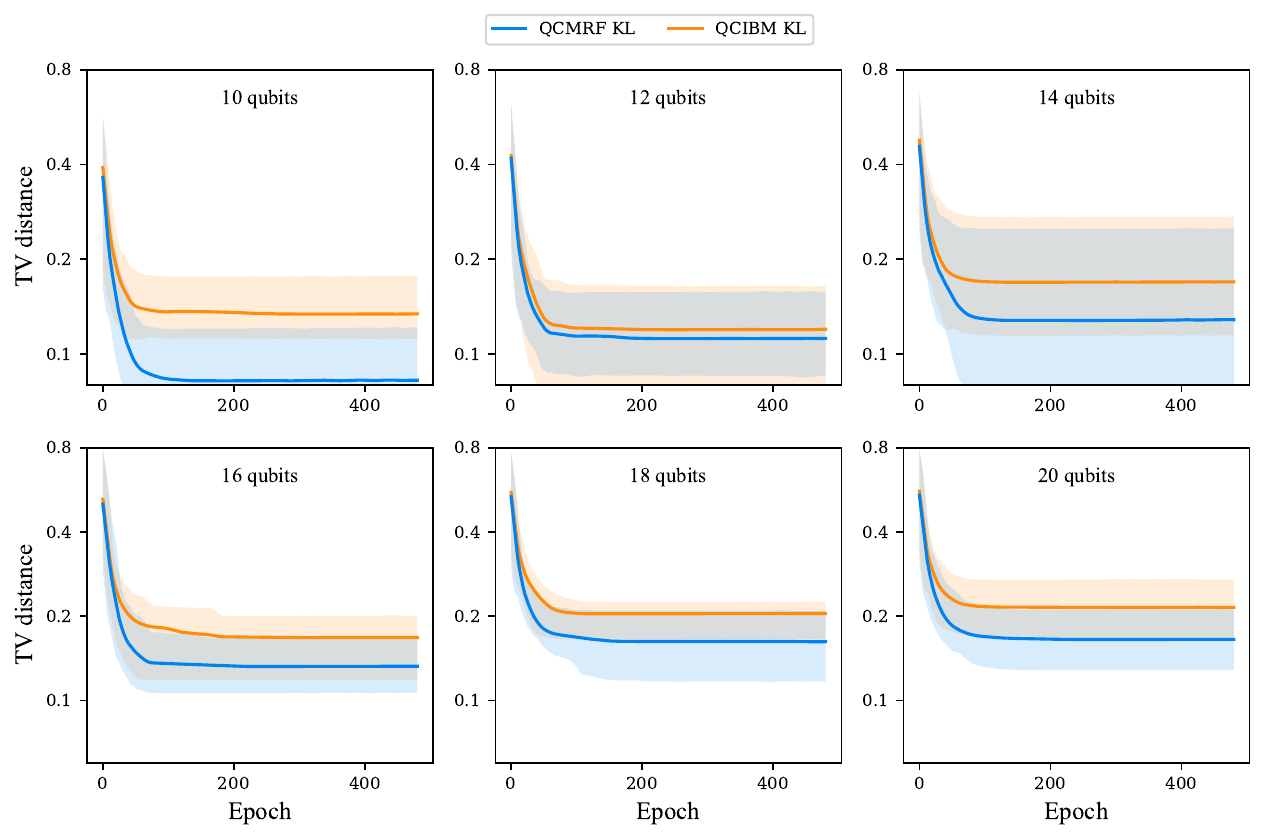}
    \caption{\textbf{QCMRF benchmark results against the problem-agnostic QCIBM using KL divergence and exact quantum circuit simulation.} These training curves show the evolution of the cost function leading to the final TV distance in Fig.~\ref{fig:scaling_kl}.}
    \label{fig:scaling_all_curves}
\end{figure*}

We further demonstrate this difference numerically on two types of graphs. In each case, we compiled the circuits to the given basis gate set using the transpilation method of Qiskit~\cite{Qiskit} with optimization level $3$. First, we explore the scaling in the number of nodes for the loop graphs shown in Fig.~\ref{fig:loop-exp}. Here we can see linear dependence in the case of BBQC and $\mathcal{O}(1)$ for QCMRF, which is expected for this very special type of graph (see Fig.~\ref{fig:depth} (left)). While these loops capture an important class, they do not provide a fair comparison, since here the triangulation of the undirected graph introduces a significant number of additional parameters for the corresponding BBQC.

As a next comparison, we concentrate on an important class of BNs, called $k$-gram models, used in natural language processing applications~\cite{manning1999foundations}. In the graph representation of a $k$-gram model, the nodes form an ordered set, in which the parents of the $l$-th node are nodes $\{l-k+1,\dots,l-1\}$, where they exist. This induces the factorization
\begin{equation}
\begin{split}
    P(X_1, \dots, X_n) & = P(X_1)P(X_2 | X_1)P(X_3 | X_1, X_2)\dots \\
    & \cdot P(X_n | X_{n-k+1}, \dots, X_{n-1}).
\end{split}
\label{eq:k-gram}
\end{equation}

These models define chordal graphs, where the number of parameters is the same in both quantum models and the maximal number of parents in the BN is the same as the size of the maximal cliques in the corresponding MN. We can observe similar scaling in $k$ for both models, as shown in Fig.~\ref{fig:depth} (right), but the depth of BBQCs is always at least an order of magnitude higher than that of QCMRF circuits. While this is not significant in a complexity theoretic perspective, it makes a significant difference when implementing on real quantum hardware, especially on near-term devices.

\section{Additional numerical simulations}
\label{appendix:numerics}

\subsection{Adversarial training}
\label{app:gan}
We numerically demonstrate how our Ansatz construction process can be integrated as a modular building block into more complex generative learning pipelines. Specifically, we consider the QGAN framework from Ref.~\cite{zoufal2019quantum}, where a quantum generator and a classical discriminator are trained adversarially. For this demonstration, we utilize the target probability distributions associated with the $8$grid graph, identical to those in Fig.~\ref{fig:double-exp} (top right).
To evaluate performance, we compare a QGAN utilizing our problem-informed QCMRF Ansatz for the generator against one utilizing the problem-agnostic QCIBM Ansatz. In both cases, the discriminator is a classical neural network composed of dense linear layers with leaky ReLU activations. The loss functions for the generator ($\mathcal{L}_G$) and discriminator ($\mathcal{L}_D$) are defined as
\begin{align}
    \mathcal{L}_G & = -\sum_{x} P_{\theta}(x) \log D(x), \\
    \mathcal{L}_D & = -\mathbb{E}_{x \sim P^{*}}[\log D(x)] - \sum_{x \in \mathcal{X}} P_{\theta}(x) \log(1 - D(x)),
\end{align}
where $D$ denotes the discriminator function. 

\begin{figure}[h]
    \centering
    \includegraphics[width=.4\textwidth]{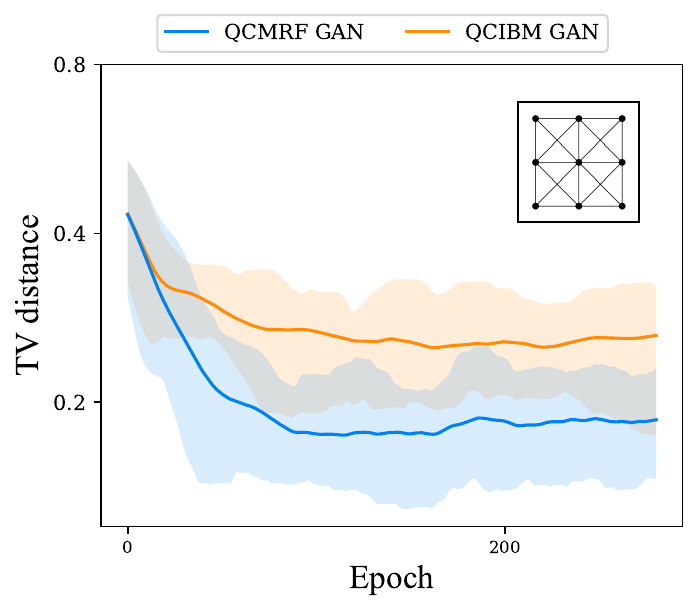}
    \caption{\textbf{Performance of QGAN models with problem-informed (QCMRF) and problem-agnostic (QCIBM) generators.} The TV distance is calculated from $10$ independent problems with the same $8$-grid structure.}
    \label{fig:gan}
\end{figure}

Throughout the trainings, we track the evolution of the TV distance. As shown in Fig.~\ref{fig:gan}, averaged across $10$ independent problem instances, the QGAN equipped with the QCMRF Ansatz consistently achieves better performance compared to the generic QCIBM baseline. This experiment provides empirical evidence that our problem-informed Ansatz construction yields tangible performance advantages when deployed within established QML architectures.

\subsection{Scaling with exact simulation}
We also present experiments based on exact simulation, i.e., exact probabilities corresponding to infinitely many shots. Here we only consider the KL divergence based on the exact model and target probability distributions. While this training is not realistic, as it requires the exact probabilities from both distributions, this way, the simulation can be scaled up to larger problems, than the ones considered in the main text.

\begin{figure}[h]
    \centering
    \includegraphics[width=.4\textwidth]{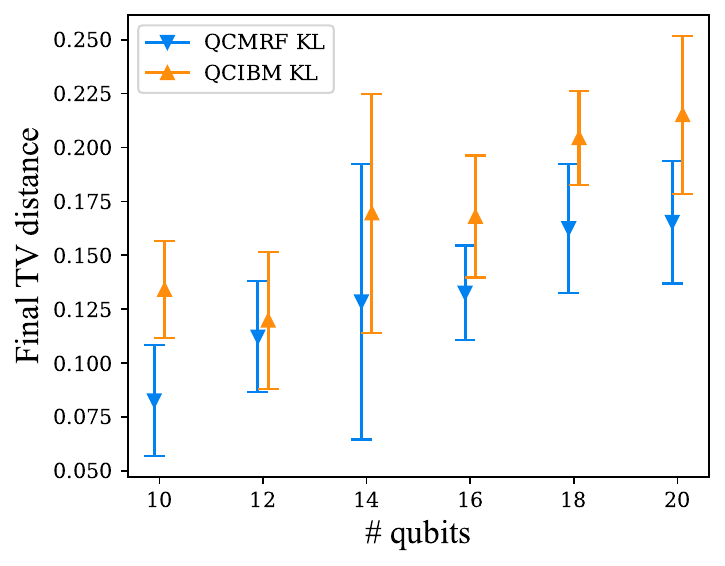}
    \caption{\textbf{Scaling of the final TV distance with the number of qubits using the KL divergence.} Here we use triangle chain graphs that correspond to $3$-gram models. The markers show the average final TV distance of $5$ independent problems and errorbars show the standard deviation.}
    \label{fig:scaling_kl}
\end{figure}

For this comparison, we use the triangle chain graph that is equivalent to the $3$-gram model, as in Eq.~\eqref{eq:k-gram}, each clique being of size $3$. In Fig.~\ref{fig:scaling_all_curves}, we present all the training curves with similar hyperparameters as in the simulation in the main text. We show the average of $5$ independent problems, having different sets of factor values.
The performance scaling with the number of qubits is shown in Fig.~\ref{fig:scaling_kl}. Here, we define the final TV distance as the average of the last $100$ training epochs in each case.
In this figure, markers show the average final TV distance among the $5$ problems, and errorbars show the standard deviation around the mean.

The separation in the average performance between the QCMRF and QCIBM is independent of the problem size, while the number of parameters is $\mathcal{O}(n)$ and $\mathcal{O}(n^2)$ respectively. For $20$ qubits, this amounts to $135$ and $270$ parameters.

\subsection{Comparison with classical methods}

\begin{figure}[h]
    \centering
    \includegraphics[width=.4\textwidth]{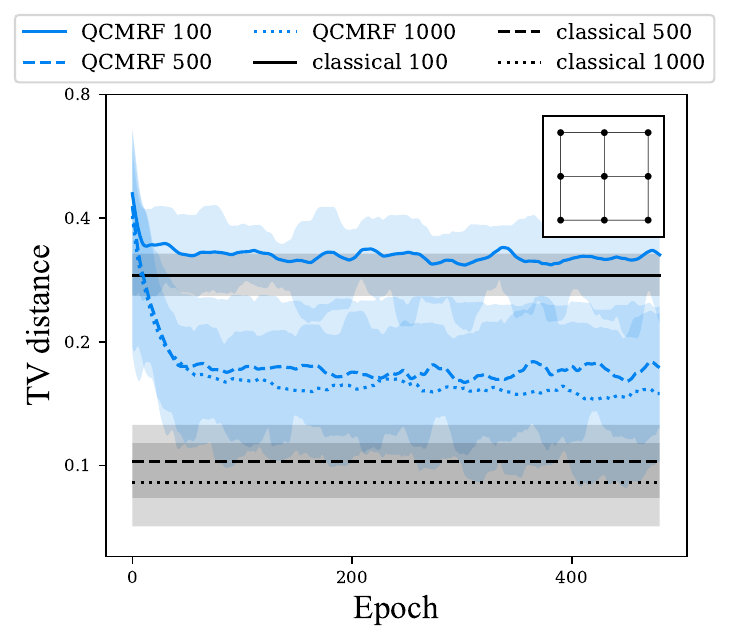}
    \caption{\textbf{Benchmark results against classical BNs.} We consider training sets of size $100$, $500$ and $1000$ for both the classical and quantum models. The QCMRFs are trained with the MMD loss function. The TV distance is computed exactly in each case.}
    \label{fig:cl_benchmark}
\end{figure}

While we concentrate on comparing QCMRF to other quantum models, here we also provide a preliminary comparison with a classical method. We consider the grid graph topology with $9$ nodes, and turn the corresponding MN into a BN. The parameters of the resulting BN are fitted to the training set using maximum likelihood estimation, as implemented in the pgmpy Python package~\cite{Ankan2024}. After fitting, the classical model to the data, we calculate the corresponding probability distribution exactly and compute the TV distance from the target distribution. We repeat this procedure for $100$, $500$ and $1000$ training samples. The corresponding results are shown in Fig.~\ref{fig:cl_benchmark}. Similarly, the QCMRFs are trained with the same number of training samples, using the MMD loss function. In each case, we run $5$ independent problems and plot the results similarly to previous experiments.

While our model shows similar performance with small training sets, its performance is weaker when the number of samples is increased. This highlights the need for more enhanced training methods for quantum models, however, this does not affect the potential of our model to have quantum advantage in sampling the learned probability distribution, since these classical methods are known to be unscalable.

\bibliographystyle{ieeetr}
\bibliography{references}

\end{document}